\documentclass[sigconf,nonacm]{acmart}
\usepackage{url}
\usepackage{multirow}
\usepackage{graphics}
\usepackage{enumitem}
\usepackage{color, colortbl}
\usepackage{lipsum}
\usepackage{tabularx,booktabs}
\usepackage[section]{placeins}
\usepackage{graphicx}
\usepackage{comment}
\usepackage[para]{footmisc}
\usepackage{rotating}
\usepackage{tablefootnote}
\restylefloat{table}
\usepackage{breakurl}
\usepackage[nobiblatex]{xurl}
\definecolor{lightgray}{rgb}{0.83, 0.83, 0.83}
\usepackage{caption}
\usepackage{subcaption}  

\DeclareCaptionLabelFormat{simplab}{Figure~#2}
\DeclareCaptionFormat{nocap}{}
\DeclareCaptionSubType*{figure}

\makeatletter
\def\@copyrightspace{\relax}
\makeatother

\makeatletter
\newcommand\notsotiny{\@setfontsize\notsotiny{5.5}{6.5}}
\makeatother

\usepackage{caption}
\copyrightyear{2023}
\acmYear{2023}
\setcopyright{acmlicensed}\acmConference[WiSec '23]{Proceedings of the 16th ACM Conference on Security and Privacy in Wireless and Mobile Networks}{May 29-June 1, 2023}{Guildford, United Kingdom}
\acmBooktitle{Proceedings of the 16th ACM Conference on Security and Privacy in Wireless and Mobile Networks (WiSec '23), May 29-June 1, 2023, Guildford, United Kingdom}
\acmPrice{15.00}
\acmDOI{10.1145/3558482.3590173}
\acmISBN{978-1-4503-9859-6/23/05}

\begin{document}


\date{}

\title{A Quantitative Study of SMS Phishing Detection}

\author{\normalsize{Daniel Timko}}
\affiliation{%
  \small
  \institution{California State University San Marcos}
   \city{San Marcos}
   \state{CA}
   \country{USA}}
\email{timko002@csusm.edu}

\author{\normalsize{Daniel Hernandez Castillo}}
\affiliation{%
  \small
  \institution{California State University San Marcos}
   \city{San Marcos}
   \state{CA}
   \country{USA}}
\email{herna1003@csusm.edu}

\author{\normalsize{Muhammad Lutfor Rahman}}
\affiliation{%
  \small
  \institution{California State University San Marcos}
   \city{San Marcos}
   \state{CA}
   \country{USA}}
\email{mlrahman@csusm.edu}

\thispagestyle{empty}

\begin{abstract}
With the booming popularity of smartphones, threats related to these devices are increasingly on the rise. Smishing, a combination of SMS (Short Message Service) and phishing has emerged as a treacherous cyber threat used by malicious actors to deceive users, aiming to steal sensitive information, money or install malware on their mobile devices. Despite the increase in smishing attacks in recent years, there are very few studies aimed at understanding the factors that contribute to a user's ability to differentiate real from fake messages.

To address this gap in knowledge, we have conducted an online survey on smishing detection with 187 participants. 
In this study, we presented them with 16 SMS screenshots and evaluated how different factors affect their decision making process in smishing detection. Next, we conducted a post-survey to garner information on the participants' security attitudes, behavior and knowledge.
Our results highlighted that attention and security behavioral scores had a significant impact on participants' accuracy in identifying smishing messages. We found that participants had more difficulty identifying real messages from fake ones, with an accuracy of 67.1\% with fake messages and 43.6\% with real messages.
Our study is crucial in developing proactive strategies to encounter and mitigate smishing attacks. By understanding what factors influence smishing detection, we aim to bolster users’ resilience against such threats and create a safer digital environment for all.

\end{abstract}

\maketitle
\section{Introduction}

\hspace{0.4cm}Phishing stands out as one of the most prevalent social engineering attacks in cybersecurity, with Cisco reporting that phishing accounts for 90\% of data breaches in the U.S.~\cite{cisco}. These attacks collectively lead to billions of dollars in losses annually~\cite{whitehouse}. Attackers are now extending their efforts into the mobile domain, introducing ransomware, spyware, or adware onto victims' mobile devices through phishing~\cite{bitdefender}. This malicious software can provide attackers access to sensitive information, including passwords, credit card details, location data, and social security numbers, resulting in significant financial damage~\cite{nagunwa2014behind}.

A survey by Openmarket reveals that 75\% of Millennials prefer texting over phone calls, and 83\% read SMS messages within 90 seconds of receipt~\cite{eztextingstats}. With the ubiquity of smartphones in daily life, attackers are targeting billions of users~\cite{statista}. RoboKiller's data indicates a surge in spam texts, with over 225 billion sent in 2022, leading to losses exceeding 22 billion dollars~\cite{robocall}. This represents a 157\% increase from the previous year and a staggering 307\% rise since 2020. According to the Federal Trade Commission, in 2022, consumers reported losing \$330 million to text message scams, which is more than double the amount reported in 2021~\cite{FTC300million}.

Numerous classical studies~\cite{dhamija2006phishing,sheng2010falls, alsharnouby2015phishing} have sought to understand what makes users vulnerable to email-based phishing attacks. However, research in the realm of smishing remains limited. In a recent study by Rahman et al.,~\cite{rahman2023users}, an empirical examination of smishing was carried out. They sent eight fake SMSes to 265 users to gauge the effectiveness of smishing attacks. Their findings revealed that up to 16.92\% of the recipients could have fallen victim to these attacks. In a separate study on smishing, Blancaflor et al.~\cite{blancaflor2021let} discovered that one in 24 targeted users clicked on the phishing URLs within the SMS. These investigations delved into the number of fake SMSes that reached participants, considering that smishing messages could be intercepted by SMS gateways, mobile carriers, or anti-smishing apps on users' devices~\cite{timko2023commercial}. 

To gain a comprehensive understanding of the issue, a more controlled experiment is needed—one that evaluates users' ability to detect smishing and examines their interactions with both real and fake messages. While previous studies~\cite{rahman2023users,blancaflor2021let} have investigated smishing from the perspective of whether a user actively becomes a victim, we intend to approach the topic from the users' perception of the messages. We explore what triggers users' suspicion that a message may be fraudulent, thus causing them to lose their comfort or willingness to interact with it. Additionally, there is nuance between a user actively interacting with a phishing message and whether they believe it is legitimate. End users may perceive a phishing message as legitimate or a legitimate message as fraudulent, while still refraining from interacting with it. By investigating this nuance, we can build a broader picture of smishing and, in turn, help focus security experts and developers on the factors causing end users to make these classification errors.

In this paper, our objective is to bridge the existing gaps by undertaking a comprehensive online study. This will offer a more profound insight into the influence of messaging content and user characteristics on interaction rates, comfort levels, and judgement between real and fake messages. In the main part of the study, we carried out an online user survey focused on smishing detection. To draw correlations with various factors, we incorporated a post-survey that encompasses several standardized metrics.

We have the following Research Questions (\textbf{RQs}) in our study. 

\begin{enumerate}
    \item \textbf{RQ1: } How do the user's likely interaction rates, comfort level, brand familiarity and messaging history influence their ability to correctly identify real and fake messages?
    \item \textbf{RQ2:} How do demographic factors affect users' interaction rates, comfort levels, and their ability in identifying real versus fake messages?
    \item \textbf{RQ3:} How do message attributes influence user interaction rates, comfort levels, and the ability to distinguish between real and fake messages?

    \item \textbf{RQ4:} How do user behavioral metrics relate to their interaction rates, comfort levels, and ability to distinguish real from fake messages?

\end{enumerate}

To address these questions, we collected data on participants' engagement levels with both real and fake messages, along with demographic details. Our experiment aimed to determine how these factors influenced participants' inclination and comfort in interacting with the messages. Additionally, we explored how these factors correlated with their past interactions with the brand. In the second round, participants were asked to determine the authenticity of the messages. Finally, we administered post-survey tests to assess participants' behavioral and attention scores. By analyzing the relationship between these scores, engagement levels, and smishing detection capabilities, we gain insights into how these factors influence a user's susceptibility to smishing attacks.

\textbf{Our Contributions:} In summary, we have made the following contributions in this study.

\begin{enumerate}[leftmargin=*]
   \item We designed and conducted an online survey with 187 participants to gain insights into how different user and message factors affect users' likelihood to interact, comfort levels with interacting and their judgement of real and fake messages.
   \item We explore a series of user demographics, attitudes and behaviors in order to identify statistically significant trends that may lead users to be more likely to interact with real and fake messages.
\end{enumerate}

\textbf{Key Results:} 
Our results reveal that participants' likelihood of interaction and comfort levels significantly predict their accuracy in correctly identifying real and fake messages. These findings underscore the importance of user perception in smishing detection tasks. Demographic factors such as ethnicity, age, income, and occupation also had a significant impact on participants' propensity to interact, their comfort levels, and their accuracy in dealing with messages. Furthermore, we observed that different elements within messages influenced accuracy differently; for fake messages, scrutinizing the sender and URL proved crucial for correct identification, whereas for real messages, the most influential aspects were the entity and the call-to-action. Additionally, we noted significantly different effects on comfort levels and interaction rates depending on which parts of the messages users focused on. The study highlights the necessity for improvements in user training and the development of SMS security features that focus user attention on key aspects of messages.

\textbf{Implications of Our Work:} We have conducted this study and unveiled both human factors and SMS attributes in smishing scenarios. We believe that our study will provide a solid foundation for designing security warnings, developing anti-smishing technologies, and creating smishing awareness programs. We have provided details on the implications of our work in the latter part of this paper.

\section{Related Works}
\label{sec:relatedworks}

\paragraph{Phishing Detection User Studies}
\hspace{0.4cm} Research by Dhamija et al. \cite{dhamija2006phishing} marked the first investigation into understanding the most effective strategies for deceiving victims in phishing attacks. The study revealed that people consistently overlooked security indicators, lacked comprehension of their functionality, and found anti-phishing web indicators to be ineffective. Following Dhamija et al.'s research, several similar user studies emerged, featuring participant mock scenarios \cite{downs2006decision, downs2007behavioral, sheng2010falls}, where individuals assumed fictional roles.

In a phishing detection user study, Alsharnouby et al. \cite{alsharnouby2015phishing} discovered that participants primarily relied on website appearance, a method proven to be flawed \cite{das2019all}. Downs et al. \cite{downs2006decision} identified that people recognized the use of social engineering in phishing attacks, but lacked an understanding of the associated risks. Neupane et al. \cite{neupane2015multi} conducted a multimodal phishing detection study using brain imaging technologies. They presented real and fake website screenshots to users, revealing that users subconsciously process real sites differently from fake sites.

Lastly, research has focused on the role demographics play in phishing detection. Wash \cite{wash2020experts} found that IT experts utilize a three stage strategy to identify phishing attacks. Baki et al.~\cite{baki2022sixteen} found that the elderly outperformed their younger counterparts in detecting fraudulent emails and websites. Similarly, Luga et al.~\cite{iuga2016baiting} found that men were more likely to successfully detect phishing attempts than women. 

\paragraph{Smishing Research}
\hspace{0.4cm}In a smishing study, Rahman et al.~\cite{rahman2023users} mentioned that even responding to messages with a request to cease communications can provide a smishing attacker with valuable information. It shows that the phone number is active and the target is willing to engage with suspicious messages. The authors witnessed participants who were willing to interact with phishing messages out of curiosity, even knowing the messages were fraudulent. This can lead to an increased chance of receiving these fraudulent messages and opening users up to future attacks. Smishing research has also focused on aspects such as the evaluation of existing technologies. Timko et al.~\cite{timko2023commercial} investigated how effective select bulk messaging services, carriers, and anti-smishing apps were at blocking smishing using a pool of 20 real and fake messages. They found current anti-smishing tools ineffective to protect against modern threats. 

The existing literature on smishing detection falls short in comprehensively addressing user perceptions. Loxdal et al. \cite{loxdal2021phishing} conducted a phishing detection user study where participants used the browser app from within a smartphone instead of a computer. They found those who focused on the URL were more likely to correctly determine legitimacy than those who focused on other aspects of the site. Yeboah-Boateng et al. \cite{yeboah2014phishing} conducted a study in which researchers asked users about their opinions and perceptions of not just smishing, but also phishing and vishing. The study found that most users had a low level of concern for the threat of smishing, despite 15\% of their participants being victims of smishing scams. 
 
\section{Background}
\label{sec:background}

\hspace{0.4cm}To measure the effect of different parts of a message on participant decisions, we defined areas of interest (AOI) in each message image and associated user data with clicks in those areas. Alsharnouby et al.'s work~\cite{alsharnouby2015phishing} explored grouping user eye tracking data on phishing websites called AOI and measuring how they affect a participant's ability to successfully identify phishing and legitimate websites. Participants' performance scores were attributed to these AOI to measure how attention to these areas affected their decision-making.

We utilized research by Rahman et al.~\cite{rahman2023users} to identify similar AOI in SMS messages. In their work, variations within message attributes, including the sender, entity, method, call to action, and scenario, were compared to determine their impact on susceptibility to smishing attacks. Here, the sender is the email, phone number, or short code that sent the message. The entity is defined as the perceived organization or brand identified in the message content as the message originator. The method is how the user is asked to respond or interact with the message. The call to action is what the message is asking the user to do. Finally, the scenario refers to the content of the message that motivates the user to take some action. Their results showed that some entities and user actions lead to significantly different rates of users falling for smishing attacks. Comparatively, in this current study we look at  between the effects of different message attributes on the success rate of smishing attempts.

\section{Methodology}
\label{sec:methodology}

\hspace{0.4cm}
In this section, we present an overview of our study methodology, encompassing ethical considerations, recruitment, demographics, and an explanation of our online survey. A total of 187 smartphone users participated in our study, which comprised of a first round and second round with multiple-choice questions. In the first round task, participants were not informed of the smishing related aspect of our study. The second round involved the smishing detection task, where participants assessed the legitimacy of SMS screenshots.

\subsection{Ethical Consideration And Mitigating Biases}

\hspace{0.4cm} To ensure the ethical conduct of our study, we obtained approval from our university's Institutional Review Board (IRB). At the survey's outset, participants were informed of potential risks through a consent form. Prior to engaging in the first round task, participants had to explicitly agree to the outlined conditions in the consent form. Participation in our study was entirely voluntary, with individuals retaining the right to withdraw consent at any point without repercussion; at withdrawal, all personally identifiable data collected was promptly destroyed. 

To mitigate bias in our study results, we provided participants with an incomplete disclosure regarding the study's nature. Participants were informed that they would be answering questions related to SMS messages without revealing the study's security-related focus. Our flyer framed the goal of our research as "Understand Mobile SMS User Behavior". This was done to mitigate the response bias, as introducing the topic of smishing may cause the participants to alter their perception of the message or respond more cautiously than in a natural setting.

\subsection{Recruitment and Demographics}
\hspace{0.4cm} Our recruitment process involved the distribution of advertising flyers through personal and professional social networks, spanning platforms including Reddit, Twitter, and LinkedIn,in addition to school email groups.
Specifically targeting smartphone users ensured familiarity with SMS messaging. Using Qualtrics as our survey host, we leveraged its features to screen out participants without US IP addresses.
As an incentive for participating in the study, we conducted an optional opportunity drawing and gave each of the 12 individuals a \$50 Amazon gift card. Participants were asked to provide an email address for contact purposes in case they won. This process resulted in a total of 187 participants recruited for our study. After recruitment, we performed a post hoc power analysis using GPower~\cite{gpower} to determine if we had sufficient power to detect medium and large effect sizes. With an alpha of .05 the results showed that we are able to achieve over .8 power for detecting medium and large effect sizes for our within group analysis. 

\subsection{Real and Fake Messages Used}

\hspace{0.4cm}The group of SMS screenshots used in the study was selected based on the rate of brand imitation through phishing. Out of the eleven brands used, five were identified as among the most imitated brands in phishing attacks \cite{checkpoint2022dhl}. Conversely, we included one brand that was not well-known, chosen from the personal network of a member of our research group. These smishing messages originated from SmishTank~\cite{smishtankweb, timko2023commercial}, where messages are collected from various phones and included different sender types, such as short codes, emails, and phone numbers. The full list of SMS used in this study can be found in our appendix in Table~\ref{fig:msglist}.

To better represent the current landscape of smishing attacks, we opted to include a screenshot of a romance scam. Romance scams pose a significant threat to users and constitute a multi-million dollar industry in the US \cite{FTC2023}. In total, the survey included nine real and seven fake screenshots. Following a precedent set by prior studies~\cite{loxdal2021phishing, dhamija2006phishing, alsharnouby2015phishing}, we used an uneven number of real and fake examples. The larger proportion of real messages is more likely to apply to real-world situations, which enhances the ecological validity of our experiment. Participants were not informed about the total number of real and fake messages in advance.

\subsection{Main Survey}
\hspace{0.4cm}The survey was conducted using Qualtrics, and participants were required to provide consent to participate and commit~\cite{commitqualtrics} to offering thoughtful answers. In the first task, participants were presented with an SMS screenshot and asked five questions about the image. This five-question process was repeated 16 times in randomized order for each participant. The complete survey procedure is illustrated in Figure~\ref{fig:mainsurvey} in the appendix. Next, we will discuss each of the questions used in our first and second round in the study.

\subsubsection{First Round Questions}
\hfill 

\textbf{Question 1 - Interaction Rate.} Initially, participants were asked to rate their likelihood of interacting with the given message on a six-point Likert Scale from `Definitely' to `Definitely not.' Interacting with the message was defined as following the URL link, replying to the text or calling the number in the message. A similar user study on phishing emails~\cite{VISHWANATH2011576} measured participants' likelihood to respond to a phishing. Previous research has highlighted that there is some nuance between a user's likelihood of interacts with messages and their belief that it is fraudulent. For example, some users in phishing studies have noted that, out of curiosity, they willingly interacted with a message even though they believed it was phishing~\cite{rahman2023users}. 

\textbf{Question 2 - Comfortability Level.} Subsequently, participants were asked, "How comfortable would you feel interacting with this SMS?" on a five-point Likert scale from `very comfortable' to `very uncomfortable,' with an option to select `prefer not to say.' This question was inspired by a previous study on security indicators on websites~\cite{Thompson2019}, where users were shown images of login pages and asked to rate their comfort with logging in. While there should be some overlap between the answers to this question and the their willingness to interact, we anticipate some differences. Participants who perceive a message as fake may paradoxically feel more comfortable interacting with it, believing they know what to look out for and won't compromise their information. 

\textbf{Question 3 - AOI Heatmap.} We integrated a heat map into our survey, instructing participants to click on message screenshots to indicate the locations affecting their comfort with interacting. This approach, aimed at identifying areas of interest to measure participant focus in security-related tasks, has been well-documented~\cite{Thompson2019, 10.1186/s13673-016-0065-2, loxdal2021phishing}. Participants were asked to click between three and five times within each message, and the click positions were recorded. These clicks enable us to compare the effects of different parts of the message, referred to as AOI, on participants. Bounding rectangles were drawn around the AOI after the survey concluded and clicks within these rectangles were marked as clicks on the AOI. Since participants were not made aware of these AOIs during the survey, clicks outside the AOI should not be interpreted as a lack of attention. 

\textbf{Question 4 - Familiarity with Brand.} Next, participants were asked a question about their familiarity with the business brand mentioned in the SMS. In a previous study, participants who were familiar with businesses mentioned in phishing emails were more likely to adopt risky behavior when attempting to accurately identify authentic phishing emails~\cite{wang2016overconfidence}. Here, we employed a similar question to measure its effect on their ability to distinguish between real and fake SMS. Their familiarity with the brand was assessed on a five-point scale ranging from 'extremely familiar' to 'not familiar at all.'

\textbf{Question 5 - Brand Messaging History.} With the fifth question, we asked participants a yes or no inquiry about whether they had received messages from the given brand in the past. This, in conjunction with the previous question, helps us understand if a user's past interaction with a brand will improve their accuracy in identifying real and fake versions of their messages.

\subsubsection{Second Round Questions}In the second round, participants were then presented with their responses to each "how likely they were to click on one of the links, reply to the text, or call the number" first round question, along with the respective screenshot. Participants were asked again, "How likely are you to interact with this SMS?" with the same options as in the first round. This was used to measure if after seeing the message a second time caused them to change their mind. Finally, we asked participants, "Do you think this SMS is real or fake?" This process repeated for all 16 messages. A similar approach has been used in prior phishing related user study tasks~\cite{neupane2015multi}.

\subsection{Post-Survey}
\hspace{0.4cm}Participants completed five post-survey questionnaires to comprehensively assess the participants' security and attention behavior. Initially, participants responded to the RSeBIS questionnaire, a revised version of the Security Behavior Intentions Scale (SeBIS) \cite{sawaya2017self, egelman2015scaling}. This questionnaire gauges user security approaches, encompassing topics like password creation and device updates. Additionally, we aimed to understand participant attitudes toward utilizing security tools, administering the Security Attitudes (SA-6) questionnaire. Comprising six questions, the SA-6 is designed to measure user security behavior and willingness to adopt security methods \cite{faklaris2019self}.

Participants also completed an Internet Users' Information Privacy Concerns (IUIPC) questionnaire to gain insight into their security behavior and perceptions of online threats~\cite{Gross2021ValidityAR}. To provide a more holistic view of individuals and their relation to cybersecurity, we included a questionnaire on common security terms, such as malware and phishing. This survey utilizes a similar approach to related works~\cite{10.5555/3563609.3563635}, and was expanded to include common terms relating to SMS security from the NCSC glossary~\cite{NSCS2022}. Participants indicated their familiarity with these terms on a five-point scale ranging from `not familiar at all' to `extremely familiar.' Recognizing that understanding these terms requires cognitive capacity, we included the Attention Control Scale (ATTC) questionnaire to measure user attention control when interacting with security concepts \cite{derryberry2002anxiety}. After the post survey, participants were debriefed about the nature of the study.

\subsection{Data Quality and Attention Checks}

\hspace{0.4cm}In the initial survey collection, we received 423 responses. To uphold data quality, we implemented various quality checks in our survey. We used the Qualtrics filtering option to limit responses to only include participants with a US GeoIP location and enabled the prevent multiple submissions option. Throughout the survey we also incorporated five attention check questions to gauge the participant attention and question comprehension. These questions encompassed two widely used forms: instructed response and instructional manipulation check (IMC)~\cite{kung2018attention, commitqualtrics}. Subsequently, we excluded participants who did not fully complete the survey or who failed more than two of the attention check questions we spread throughout the questionnaires. Following these quality checks, we retained 187 participants for our analysis.

\section{Results}
\label{sec:results}

\begin{figure}[ht]
    \centering
    \includegraphics[clip, trim=0cm 0cm -1.6cm 0cm,width=\linewidth]{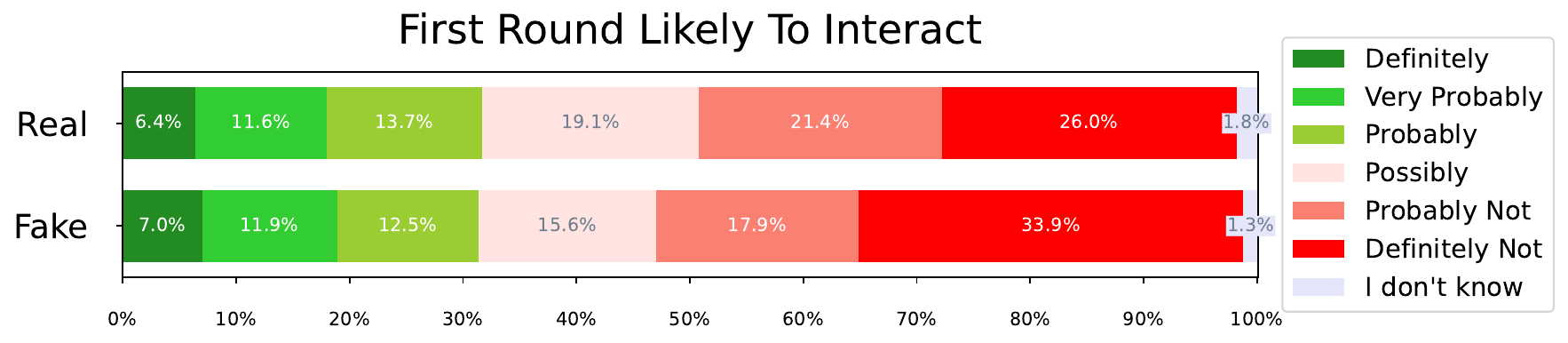}
    \includegraphics[clip, trim=0cm 0cm -1.6cm 0cm,width=\linewidth]{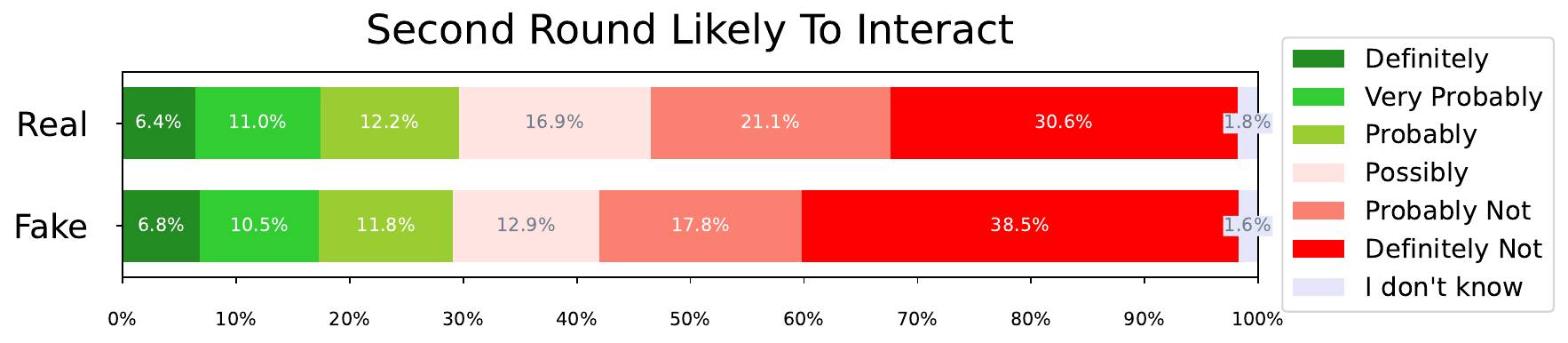}
    \includegraphics[width=\linewidth]{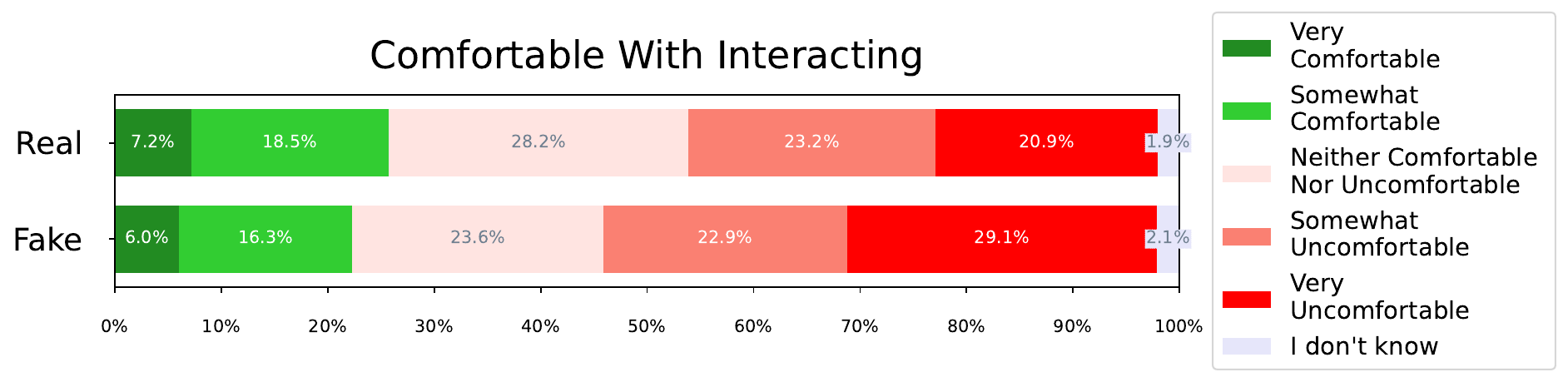}
    \includegraphics[clip, trim=0cm 0cm -.6cm 0cm,width=\linewidth]{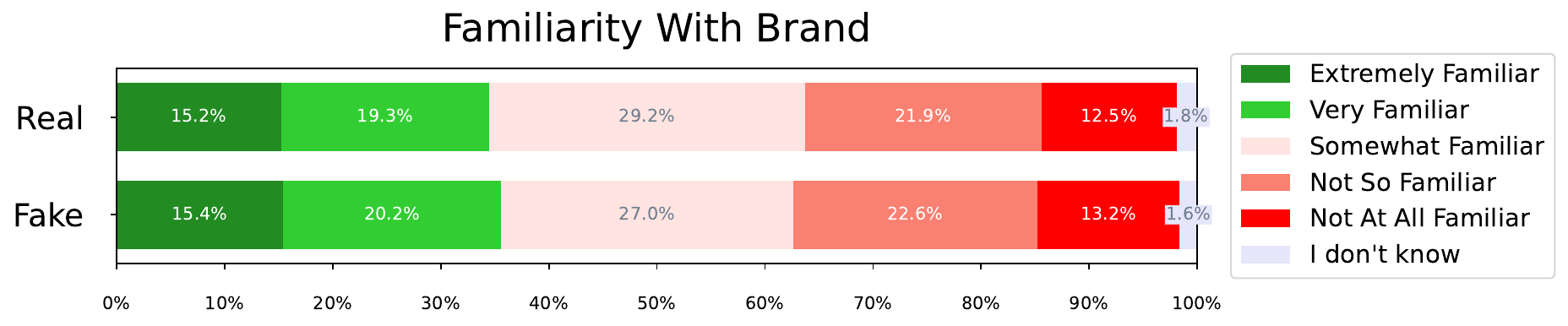}
    \includegraphics[clip, trim=0cm 0cm -2.6cm 0cm,width=\linewidth]{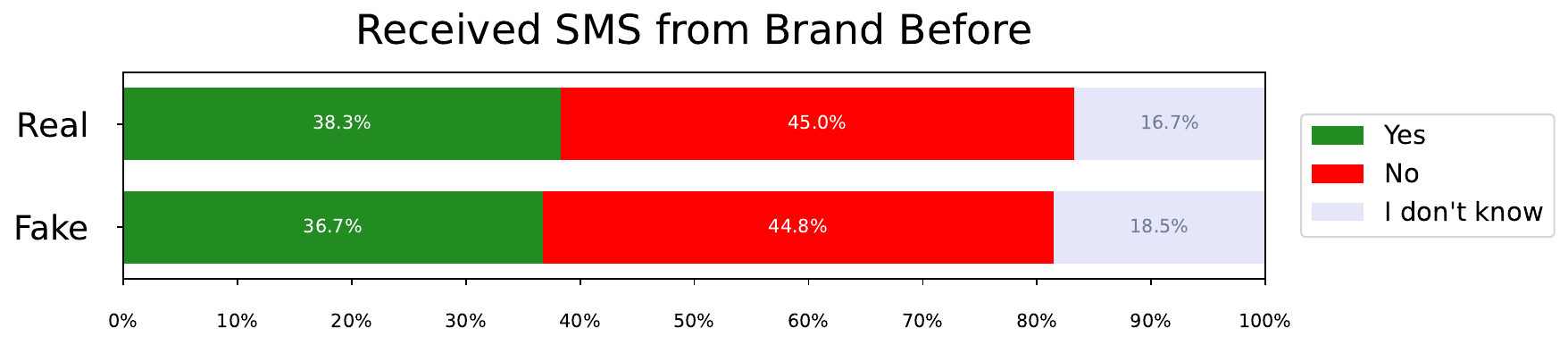}
    \includegraphics[clip, trim=0cm 0cm -2.6cm 0cm,width=\linewidth]{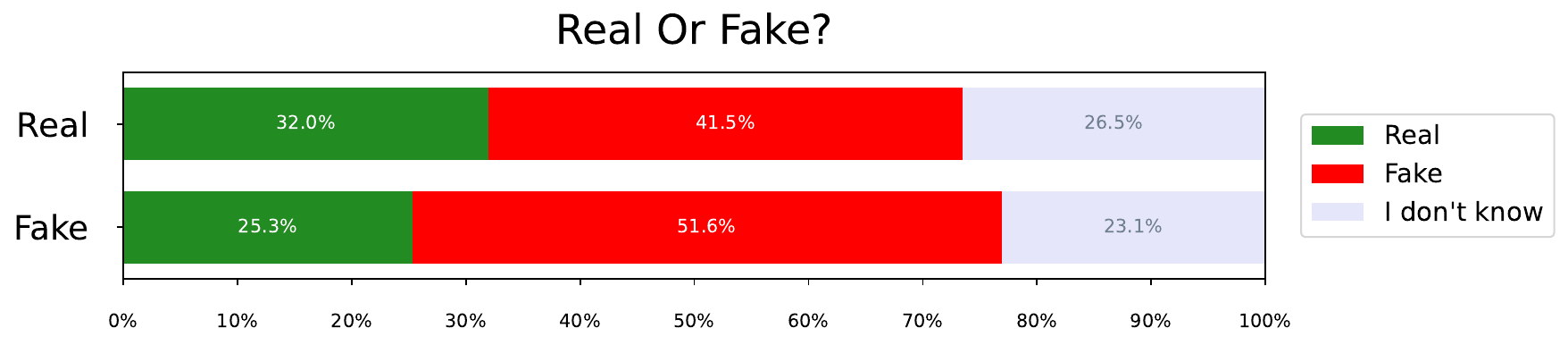}
    \caption{Breakdown of average metrics for handling each message, separated by real and fake messages.}
    \label{fig:metricsview}
    \vspace{-3mm}
\end{figure}

\subsection{Answering RQ:1}

\hspace{0.4cm}Initially, we analyzed the accuracy rates of the messages, as illustrated in Figure~\ref{tab:accuracy-rates-bymessage} in the appendix. This table presents the average rates at which participants correctly identified our messages as real or fake. The accuracy rates per message, as well as the overall accuracy for both real and fake messages, are depicted. The average accuracy per fake message ranged from 63.1\% to 73.5\%, with an overall average accuracy of 67.1\%. For real messages, our accuracy ranged from 34.0\% to 51.8\%, with an average of 43.6\%.

Next, we assessed how a user's ability to accurately distinguish between real and fake messages was influenced by metrics related to the 16 questions in the smishing detection round. The average rates for each response across the first and second rounds are depicted in Figure~\ref{fig:metricsview}. Participants were asked twice regarding their likelihood to interact with the messages. A comparison of interaction likelihood between rounds one and two revealed a decrease of 0.12 points for fake messages and 0.11 points for real messages on the six-point Likert scale. Nevertheless, when examining the mean scores of these 187 participants, we found no significant difference in the likelihood to interact with fake messages (t(370) = 0.64, \textit{p} = 0.425) or real messages (t(369) = 1.54, \textit{p} = 0.216).

In Figure~\ref{fig:metricsview}, we also present the frequency of each answer for the first round of questions with the SMS stimuli and our second round questions, categorized by real and fake messages. The mean Likert scores for the likelihood of interacting with real messages were 2.83, while the scores for fake messages were 2.70. Overall, this indicates that participants were slightly more inclined to interact with real messages than fake ones. The same trend holds true for comfortability with interacting with the messages, with a 2.67 average Likert score for real messages and 2.47 for fake messages. Participants scored brand familiarity with an average of 3.02 for fake and real messages, indicating a similar level of familiarity with both. Similarly, participants reported having received messages from the brands listed in the messages an average 55\% of the time for both real and fake messages. Lastly, we present the real and fake responses to the smishing detection task, revealing numerous false positive and false negative results. Notably, among the last two questions, a significant percentage of `I don't know' responses indicate uncertainty about whether participants had received a message before and whether the messages were real or fake.

Subsequently, we conducted a linear regression analysis on the interaction metrics described in Figure~\ref{fig:metricsview} to determine whether the metrics from our first-round smishing detection task significantly predicted participants' accuracy in identifying the message as real or fake. For each case, we tested the null hypothesis with the F-test for overall significance.

The linear regression for fake messages was statistically significant ($R^2$ = 0.518, F(4, 166) = 46.73, \textit{p} < 0.001). The regression analysis revealed that the likelihood of interacting with fake messages significantly predicted accuracy with fake messages ($\beta$ = -0.679, \textit{p} < 0.001). However, familiarity with the brand, previous receipt of an SMS from the brand, and comfortability with interacting did not significantly predict smishing detection accuracy for fake messages.

Similarly, for real messages, the linear regression was statistically significant ($R^2$ = 0.374, F(4, 164) = 28.12, \textit{p} < 0.001). The regression analysis demonstrated that the likelihood of interacting with real messages significantly predicted accuracy with real messages ($\beta$ = 0.387, \textit{p} = 0.002). Additionally, comfortability with interacting also significantly predicted accuracy with real messages ($\beta$ = 0.251, \textit{p} = 0.037). However, familiarity with the brand and previous receipt of an SMS from the brand did not significantly predict smishing detection accuracy for real messages.
Alternatively, we did not find any significant differences in regards to messaging history or familiarity with the brand.


\textbf{RQ1 Summary:}  These results highlight the correlation between interaction rates, comfort with interaction, and accuracy rates. Specifically, participants less inclined to engage with fake messages were more adept at correctly identifying them as such. Furthermore, interaction rates and comfort levels were significant predictors of accurately identifying real messages. In contrast, brand familiarity and a history of receiving messages from the brand did not significantly influence accuracy for either real or fake messages. These findings imply that while comfort with interaction may affect the likelihood of engagement, it does not impact the accuracy of identifying fake messages as much as participants' self-reported likelihood of interaction does.

\begin{table}[]
\tiny
\centering
\caption{ANOVA results for accuracy, likely to interact, and comfort rates for real and fake messages.}
\label{tab:ANOVAEffectResults}
\begin{tabular}{l|cccccccc}
\hline
 & \multicolumn{4}{c|}{\textbf{Fake Messages}} & \multicolumn{4}{c}{\textbf{Real Messages}} \\ \hline
\textbf{IV} & \textbf{F} & \textit{\textbf{p}} & \textbf{$\eta^{2}$} & \multicolumn{1}{c|}{\textbf{95\% CI}} & \textbf{F} & \textit{\textbf{p}} & \textbf{$\eta^{2}$} & \textbf{95\% CI} \\ \hline
 & \multicolumn{8}{c}{\textbf{Accuracy Rates}} \\ \hline
\textbf{Sex} & 3.39 & \textbf{.036} & .038 & \multicolumn{1}{c|}{{[}.00-.10{]}} & 0.97 & .382 & .011 & {[}.00-.05{]} \\
\textbf{Ethnicity} & 2.38 & \textbf{.031} & .079 & \multicolumn{1}{c|}{{[}.00-.14{]}} & 1.81 & .099 & .062 & {[}.00-.11{]} \\
\textbf{Age} & 2.32 & .059 & .052 & \multicolumn{1}{c|}{{[}.00-.11{]}} & 1.98 & .099 & .046 & {[}.00-.10{]} \\
\textbf{Income} & 2.23 & \textbf{.034} & .086 & \multicolumn{1}{c|}{{[}.00-.14{]}} & 1.53 & .159 & .062 & {[}.00-.11{]} \\
\textbf{Smartphone Use} & 0.17 & .955 & .004 & \multicolumn{1}{c|}{{[}.00-.01{]}} & 0.41 & .800 & .010 & {[}.00-.03{]} \\
\textbf{Education} & .638 & .745 & .030 & \multicolumn{1}{c|}{{[}.00-.04{]}} & 1.10 & .362 & .045 & {[}.00-.08{]} \\
\textbf{Occupation} & 1.57 & .113 & .097 & \multicolumn{1}{c|}{{[}.00-.13{]}} & 1.26 & .253 & .080 & {[}.00-.11{]} \\ \hline
 & \multicolumn{8}{c}{\textbf{Likelihood to Interact}} \\ \hline
\textbf{Sex} & 3.62 & \textbf{.029} & .038 & \multicolumn{1}{c|}{{[}.00-.10{]}} & 1.70 & .186 & .018 & {[}.00-.07{]} \\
\textbf{Ethnicity} & 4.60 & \textbf{\textless{}.001} & .133 & \multicolumn{1}{c|}{{[}.03-.20{]}} & 3.08 & \textbf{.007} & .094 & {[}.01-.15{]} \\
\textbf{Age} & 5.66 & \textbf{\textless{}.001} & .111 & \multicolumn{1}{c|}{{[}.03-.18{]}} & 3.98 & \textbf{.004} & .081 & {[}.01-.15{]} \\
\textbf{Income} & 5.60 & \textbf{\textless{}.001} & .180 & \multicolumn{1}{c|}{{[}.06-.25{]}} & 3.62 & \textbf{.001} & .125 & {[}.02-.19{]} \\
\textbf{Smartphone Use} & 0.57 & .687 & .012 & \multicolumn{1}{c|}{{[}.00-.04{]}} & 0.51 & .726 & .011 & {[}.00-.03{]} \\
\textbf{Education} & 1.13 & .345 & .048 & \multicolumn{1}{c|}{{[}.00-.08{]}} & 1.60 & .127 & .068 & {[}.00-.11{]} \\
\textbf{Occupation} & 2.17 & \textbf{.018} & .120 & \multicolumn{1}{c|}{{[}.00-.16{]}} & 0.68 & .752 & .041 & {[}.00-.05{]} \\ \hline
 & \multicolumn{8}{c}{\textbf{Comfort with Interacting}} \\ \hline
\textbf{Sex} & 3.63 & \textbf{.029} & .038 & \multicolumn{1}{c|}{{[}.00-.10{]}} & 0.65 & .525 & .007 & {[}.00-.04{]} \\
\textbf{Ethnicity} & 6.38 & \textbf{\textless{}.001} & .176 & \multicolumn{1}{c|}{{[}.06-.25{]}} & 4.18 & \textbf{\textless{}.001} & .123 & {[}.03-.19{]} \\
\textbf{Age} & 5.43 & \textbf{\textless{}.001} & .107 & \multicolumn{1}{c|}{{[}.02-.18{]}} & 3.34 & \textbf{.012} & .069 & {[}.00-.13{]} \\
\textbf{Income} & 5.57 & \textbf{\textless{}.001} & .180 & \multicolumn{1}{c|}{{[}.06-.25{]}} & 4.80 & \textbf{\textless{}.001} & .159 & {[}.05-.23{]} \\
\textbf{Smartphone Use} & 0.25 & .911 & .005 & \multicolumn{1}{c|}{{[}.00-.02{]}} & 0.48 & .750 & .011 & {[}.00-.03{]} \\
\textbf{Education} & 0.81 & .597 & .035 & \multicolumn{1}{c|}{{[}.00-.06{]}} & 0.99 & .444 & .043 & {[}.00-.07{]} \\
\textbf{Occupation} & 2.35 & \textbf{.010} & .129 & \multicolumn{1}{c|}{{[}.01-.17{]}} & 0.72 & .723 & .043 & {[}.00-.05{]} \\ \hline
\end{tabular}
\end{table}

\subsection{Answering RQ:2}

\hspace{0.4cm}To investigate the effect of demographics on the behavioral predictors that influence users’ susceptibility to smishing attacks, we conducted an ANOVA test. This test aimed to compare the impacts of various demographic factors on the likelihood of users to interact with both real and fake messages, as well as their accuracy in identifying messages. The ANOVA test results, along with their effect sizes, are presented in Table~\ref{tab:ANOVAEffectResults}. For a detailed breakdown of our demographics, please refer to Table~\ref{tab:DemographicAttribute2} in the appendix. Given that we do not assume homogeneity of variances in our responses, we employed a Games-Howell post hoc test for multiple comparisons when the results of our ANOVA test were deemed significant. The significance of the results of this test are to inform the understanding of the users' vulnerabilities to smishing attacks in different populations. Furthermore, by identifying their significant differences in susceptibility to smishing attacks, these findings can allow for targeted future research and resource allocation, directing efforts towards populations most vulnerable.

\vspace{-2mm}

\subsubsection{Sex}
\hfill

\textbf{Likely to interact}: The post hoc test results for average accuracy when identifying fake messages were not significant.

\textbf{Comfortability with interacting}: 
The post hoc test results show that the average comfort levels with fake messages was significantly lower for female participants in comparison with male participants(\textit{p} = 0.030, 95\% C.I. = [-0.66, -0.03]).

\textbf{Accuracy}: 
The post hoc test results show that the average accuracy with fake messages was significantly higher for female participants in comparison with male participants(\textit{p} = 0.032, 95\% C.I. = [0.01, 0.26]).

\subsubsection{Ethnicity}
\hfill

\textbf{Likely to interact}: 
The post hoc test results show that the average likely interaction rates with fake messages was significantly lower for Hispanic participants in comparison with White (\textit{p} $<$ 0.001, 95\% C.I. = [-2.06, -0.87]), Black (\textit{p} = 0.001, 95\% C.I. = [-2.91, -0.59]) and Native American (\textit{p} $<$ 0.001, 95\% C.I. = [-2.57, -0.80]). The post hoc tests results show that the average likely interaction rates with real messages was significantly lower for Hispanic participants in comparison with White (\textit{p} = 0.028, 95\% C.I. = [-2.06, -0.10]), Black (\textit{p} = 0.010, 95\% C.I. = [-2.81, -0.27]) and Native American (\textit{p} = 0.016, 95\% C.I. = [-2.38, -0.18])

\textbf{Comfortability with interacting}: 
The post hoc test results show that the average comfort level with interacting with fake messages was significantly lower for Hispanic participants in comparison with White (\textit{p} $<$ 0.001, 95\% C.I. = [-1.75, -0.91]), Asian (\textit{p} = 0.019, 95\% C.I. = [-2.05, -0.14]), Black (\textit{p} = 0.003, 95\% C.I. = [-2.24, -0.39]) and Native American (\textit{p} $<$ 0.001, 95\% C.I. = [-2.06, -0.83]). The post hoc test results also show that the average comfort level with interacting with real messages was significantly lower for Hispanic participants in comparison with White (\textit{p} = 0.033, 95\% C.I. [-2.04, -0.07]) and Native American (\textit{p} = 0.017, 95\% C.I. = [-2.41, -0.17]).

\textbf{Accuracy}:  
The post hoc test results show the average accuracy with fake messages was significantly higher for Hispanic participants in comparison with White (\textit{p} $<$ 0.001, 95\% C.I. = [0.15, 0.43]), Black (\textit{p} = 0.034, 95\% C.I. = [0.02, 0.76]) and Native American (\textit{p} = 0.011, 95\% C.I. = [0.07, 0.73]) participants.

\subsubsection{Age}
\hfill

\textbf{Likely to interact}: 
The post hoc test results show that the average likely interaction rates with fake messages was significantly lower for participants in the 18-24 age category than those in the 25-34 (\textit{p} = 0.034, 95\% C.I. = [-1.36, -0.04]) and 35-44 (\textit{p} $<$ 0.001, 95\% C.I. = [-1.87, -0.40]) category. The post hoc test results show that average likely interaction rates with real messages were not significant.

\textbf{Comfortability with interacting}: 
The post hoc test results show that the average comfort levels with fake messages was significantly lower for participants in the 18-24 age category than those in the 35-44 (\textit{p} = 0.002, 95\% C.I. = [-1.49, -0.26]) and 45-54 (\textit{p} = 0.004, 95\% C.I. = [-1.50, -0.23]) age category. The post hoc test results also show that the average comfort levels with real messages was significantly higher in participants in the 45-54 age category than those in the 18-24 (\textit{p} = 0.026, 95\% C.I. = [0.05, 1.15]) and 25-34 (\textit{p} $<$ 0.001, 95\% C.I. = [0.20, 0.62]) age category.

\textbf{Accuracy}: We found no statistically significant difference when considering the criteria of accuracy.

\subsubsection{Income level}
\hfill

\textbf{Likely to interact}:  
The post hoc test results show that the average likely interaction rates with fake messages was significantly lower for participants with an income less than \$10,000 than participants in the \$20,000-\$39,999 (\textit{p} $<$ 0.001, 95\% C.I. = [-3.08, -1.02]), \$40,000-\$59,999 (\textit{p} $<$ 0.001, 95\% C.I. = [-2.69, -1.32]) and \$80,000-\$99,999 (\textit{p} $<$ 0.001, 95\% C.I. = [-2.11, -0.71]) range. The post hoc results for average likely interaction rates with real messages were not significant.

\textbf{Comfortability with interacting}: 
The post hoc test results show that the average comfort levels with fake messages was significantly lower for participants with an income less than \$10,000 than participants in the \$20,000-\$39,999 (\textit{p} = 0.038, 95\% C.I. = [-2.08, -0.05]), \$40,000-\$59,999 (\textit{p} = 0.027, 95\% C.I. = [-2.04, -0.13]) and \$80,000-\$99,999 (\textit{p} = 0.039, 95\% C.I. = [-2.00.-0.09]) range. The post hoc results for average likely interaction rates with real messages were not significant.

\textbf{Accuracy}:  
The post hoc test results show that the average accuracy rates fake messages was significantly higher for participants with an income less than \$10,000 than participants in the \$20,000-\$39,999 (\textit{p} = 0.023, 95\% C.I. = [0.04, 0.72]), \$40,000-\$59,999 (\textit{p} $<$ 0.001, 95\% C.I. = [0.14, 0.57]) and \$80,000-\$99,999 (\textit{p} $<$ 0.001, 95\% C.I. = [0.19, 0.66]) range.

\subsubsection{Smartphone Use, Education Levels} We found no statistically significant difference when considering these criteria.

\subsubsection{Occupation}
\hfill

\textbf{Likely to interact}:  
The post hoc test results show that the average likely interaction rates with fake messages was significantly lower for students when compared to business, management, or financial (\textit{p} 0 0.004, 95\% C.I. = [-2.32, -0.29]) and IT (\textit{p} = 0.008, 95\% C.I. = [-2.70, -0.25]). 

\textbf{Comfortability with interacting}: 
The post hoc test results show that the average comfort levels with fake messages was significantly lower for students when compared to IT (\textit{p} = 0.009, 95\% C.I. = [-1.98, -0.17]), and service occupations (\textit{p} = 0.031, 95\% C.I. = [-2.15,-0.06]). 

\textbf{Accuracy};
We found no statistically significant difference when
considering the criteria of accuracy.

\textbf{RQ2 Summary:} We found that differences in the Sex, Ethnicity, Age, Income and Occupation of participants had a significant impact on the user's likely interaction rates, comfort levels and accuracy in identifying the messages. 

\begin{table}[]
\vspace{-1mm}
\centering
\scriptsize
\setlength{\tabcolsep}{4pt} 
\renewcommand{\arraystretch}{1.2} 
\begin{tabular}{llccccc}
\hline
\textbf{Category} & \textbf{Predictors} & \textbf{\#Clicked} & \textbf{$\beta$} & \textbf{Sig} & \textbf{Exp($\beta$)} & \textbf{95\% C.I.} \\ \hline
\multirow{6}{*}{\rotatebox[origin=c]{90}{Real Messages}} & Sender & 304 & -.014 & .925 & .986 & {[}.74 - 1.31{]} \\
 & Entity & 294 & .216 & .117 & 1.241 & {[}.95 - 1.62{]} \\
 & URL & 1316 & -.050 & .646 & .951 & {[}.77 - 1.18{]} \\
 & CTA & 378 & .395 & \textbf{.002} & 1.484 & {[}1.16 - 1.90{]} \\
 & Scenario & 573 & .101 & .379 & 1.106 & {[}.88 - 1.39{]} \\
 & Const. & - & -.875 & \textbf{\textless{}.001} & .417 & - \\ \hline
\multirow{6}{*}{\rotatebox[origin=c]{90}{Fake Messages}} & Sender & 239 & .671 & \textbf{\textless{}.001} & 1.956 & {[}1.42 - 2.69{]} \\
 & Entity & 278 & -.043 & .770 & .958 & {[}.72 - 1.28{]} \\
 & URL & 772 & .408 & \textbf{\textless{}.001} & 1.504 & {[}1.18 - 1.92{]} \\
 & CTA & 165 & .104 & .567 & 1.110 & {[}.78 - 1.58{]} \\
 & Scenario & 579 & -.103 & .412 & .902 & {[}.70 - 1.15{]} \\
 & Const. & - & -.249 & \textbf{.025} & .779 & - \\ \hline
\multirow{6}{*}{\rotatebox[origin=c]{90}{All Messages}} & Sender & 543 & .334 & \textbf{.001} & 1.396 & {[}1.14 - 1.70{]} \\
 & Entity & 572 & .143 & .145 & 1.154 & {[}.95 - 1.40{]} \\
 & URL & 2088 & .102 & .200 & 1.107 & {[}.95 - 1.29{]} \\
 & CTA & 543 & .218 & \textbf{.031} & 1.244 & {[}1.02 - 1.52{]} \\
 & Scenario & 1152 & .068 & .405 & 1.071 & {[}.91 - 1.26{]} \\
 & Const. & - & -.622 & \textbf{\textless{}.001} & .537 & - \\ \hline
\end{tabular}
\caption{Binary Logistic Regression results for correctly identifying real, fake, and all messages by AOI.}
\label{tab:BLR_AOI}
\end{table}

\subsection{Answering RQ:3}
\subsubsection{Areas of Interest and their relationship with accurately identifying real and fake messages}

\hspace{0.4cm}To explore the impact of message attributes on a user's likelihood to interact, comfort levels, and accuracy in identifying real and fake messages, we implemented a task where participants clicked on pictures of SMS messages to indicate which part of the image influenced their feelings about interacting. Participant clicks were then categorized into five distinct AOI: the sender, entity mentioned in the message, the URL contained in the message, the call to action in the message, and the text describing the scenario of the message. An exception was made for one of our fake messages that lacked a URL, entity, call to action, or scenario, leading to its exclusion from the AOI analysis. In total, participants generated 8,832 clicks across 187 participants and 16 images, with 4,898 of these clicks landing within one of the specified AOIs. 706 clicks resulted in an "i don't know" decision, which we separate from this analysis. The remaining 4192 clicks lead to a 'real' or 'fake' decision. In 81\% of the individual message tasks, participant clicked within at least one AOI. The messages with their corresponding AOI and user clicks can be found in Figure \ref{fig:aoi_images} in the appendix.


To assess the influence of these AOIs on the participants' ability to correctly identify real and fake messages, we employed binary logistic regression. The results of this analysis are presented in Table~\ref{tab:BLR_AOI}.

\textbf{Real Messages} Clicking on the CTA increased the odds of correctly identifying real messages by 48.4\% (95\% C.I. [1.16, 1.90])).

\textbf{Fake Messages} Clicking on the sender increased the odds of correctly identifying fake messages by 95.6\% (95\% C.I. [1.42, 2.69]), while clicking on the URL increased the odds by 50.4\% (95\% C.I. [1.18, 1.92]).

\textbf{All Messages}  Clicking on the sender increased the odds of correctly identifying all messages by 39.6\% (95\% C.I. [1.14, 1.70]) while clicking on the CTA increased the odds of correctly identifying messages by 24.4\% (95\% C.I. [1.02, 1.52]).

\begin{table}[]
\scriptsize
\begin{tabular}{llcccccc}
\hline
 & \multicolumn{1}{c}{} & \multicolumn{3}{c}{\textbf{Comfort Levels}} & \multicolumn{3}{c}{\textbf{Likely Interaction}} \\ \hline
\textbf{} & \multicolumn{1}{c}{\textbf{Predictors}} & \textbf{$\beta$} & \textbf{Sig} & \multicolumn{1}{c|}{\textbf{95\% C.I.}} & \textbf{$\beta$} & \textbf{Sig} & \textbf{95\% C.I.} \\ \hline
\multirow{6}{*}{\rotatebox[origin=c]{90}{Real Meassages}} & Sender & -.29 & \textbf{\textless{}.001} & \multicolumn{1}{c|}{{[}-.45,-.14{]}} & -0.28 & \textbf{.01} & {[}-.48,-.07{]} \\
 & Entity & .07 & .404 & \multicolumn{1}{c|}{{[}-.09,.22{]}} & 0.08 & .43 & {[}-.12,.28{]} \\
 & URL & -.04 & .531 & \multicolumn{1}{c|}{{[}-.16,.08{]}} & 0.02 & .79 & {[}-.13,.18{]} \\
 & CTA & .15 & .050 & \multicolumn{1}{c|}{{[}.00,.29{]}} & 0.18 & .06 & {[}-.01,.37{]} \\
 & Scenario & .12 & .065 & \multicolumn{1}{c|}{{[}-.01,.25{]}} & .10 & .25 & {[}-.07,.26{]} \\
 & Constant & 2.67 & \textbf{\textless{}.001} & \multicolumn{1}{c|}{{[}2.56,2.77{]}} & 2.78 & \textbf{\textless{}.001} & {[}2.64,2.91{]} \\ \hline
\multirow{6}{*}{\rotatebox[origin=c]{90}{Fake Messages}} & Sender & -.45 & \textbf{\textless{}.001} & \multicolumn{1}{c|}{{[}-.63,-.26{]}} & -.56 & \textbf{\textless{}.001} & {[}-.80,-.32{]} \\
 & Entity & -.03 & .763 & \multicolumn{1}{c|}{{[}-.20,.14{]}} & .00 & .969 & {[}-.22,.23{]} \\
 & URL & -.33 & \textbf{\textless{}.001} & \multicolumn{1}{c|}{{[}-.47,-.18{]}} & -.34 & \textbf{\textless{}.001} & {[}-.53,-.14{]} \\
 & CTA & -.09 & .395 & \multicolumn{1}{c|}{{[}-.30,.12{]}} & .03 & .841 & {[}-.25,.31{]} \\
 & Scenario & .05 & .478 & \multicolumn{1}{c|}{{[}-.09,.20{]}} & .17 & .085 & {[}-.02,.37{]} \\
 & Constant & 2.71 & \textbf{\textless{}.001} & \multicolumn{1}{c|}{{[}2.58,2.85{]}} & 2.94 & \textbf{\textless{}.001} & {[}2.77,3.12{]} \\ \hline
\end{tabular}
\caption{Linear Regression results of AOI effects on Comfort Levels and Likely Interaction Rates in real and fake messages.}
\label{tab:mlr-AOItable}
\end{table}
\subsubsection{The Effect of Area of Interest on Comfortability and Interaction Rates.}

Finally, we measured which AOI had a significant effect on the interaction rates and comfort levels across real and fake messages. The purpose was to identify which areas in real messages made participants more likely to interact and feel comfortable doing so, while determining the opposite reaction for fake messages. To assess this effect, we employed linear regression with the 5 AOIs as our predictors. This method enables us to measure the influence of each area while controlling for the contribution of others. The results of this analysis are presented in Table~\ref{tab:mlr-AOItable}.

\textbf{Interaction Rates With Real Messages}
Clicking on the sender decreased the average likelihood of interacting with a real message by 0.28 points(95\% C.I. [-.48,-.07]).

\textbf{Interaction Rates With Fake Messages}
Clicking on the sender decreased the likelihood of interacting with a fake message by 0.56
points(95\% C.I. [-.79,-.33]). Clicking on the URL also decreased the average likelihood of interacting with a fake message by 0.34 points(95\% C.I. [-.53,-.14]).

\textbf{Comfort Levels With Real Messages}
Clicking on the sender decreased the average comfort level interacting with a real message by 0.29 points(95\% C.I. [-.45,-.14]).

\textbf{Comfort Levels With Fake Messages}
Clicking on the sender decreased the likelihood of interacting with a fake message by 0.45 points(95\% C.I. [-.63,-.26]). Clicking on the URL also decreased the likelihood of interacting with a fake message by 0.33 points(95\% C.I. [-.47,-.18]).

\begin{table*}[h]
\centering
\scriptsize
\begin{tabular}{lcccccccccc}
\hline
 & \multicolumn{5}{c|}{\textbf{Real SMS}} & \multicolumn{5}{c}{\textbf{Fake SMS}} \\ \cline{2-11} 
 & \textbf{Sender \%} & \textbf{Entity \%} & \textbf{URL \%} & \textbf{Call to Ac. \%} & \multicolumn{1}{c|}{\textbf{Scenario \%}} & \textbf{Sender \%} & \textbf{Entity \%} & \textbf{URL \%} & \textbf{Call to Ac. \%} & \textbf{Scenario \%} \\ \hline
\multicolumn{11}{c}{\textbf{Comfortability of Interacting (Correct/Incorrect)}} \\ \hline
\textbf{Very Comf (C/I)} & 16.5/2.1 & 21.6/3.3 & 19.2/2.3 & 20.0/4.9 & \multicolumn{1}{c|}{20.6/3.8} & 0.0/8.5 & 0.0/12.8 & 1.8/7.6 & 1.3/7.8 & 2.3/12.2 \\
\textbf{Somewhat Comf (C/I)} & 35.3/9.4 & 35.3/12.5 & 35.5/10.0 & 35.6/11.8 & \multicolumn{1}{c|}{33.3/12.3} & 6.2/28.2 & 10.9/21.4 & 7.3/19.6 & 6.4/29.7 & 8.9/29.4 \\
\textbf{Neither (C/I)} & 17.6/18.3 & 18.6/28.8 & 26.7/30.0 & 22.2/30.5 & \multicolumn{1}{c|}{21.8/31.9} & 11.6/18.3 & 14.7/35 & 14.3/41.2 & 12.8/32.8 & 10.7/32.1 \\
\textbf{Somewhat Uncomf (C/I)} & 22.4/28.3 & 14.7/22.3 & 13.4/25.5 & 17.0/21.7 & \multicolumn{1}{c|}{18.2/25.2} & 19.9/31 & 18.6/19.7 & 21.6/20.4 & 24.4/23.4 & 27.6/19.0 \\
\textbf{Very Uncomf (C/I)} & 8.2/41.9 & 9.8/33.2 & 5.2/32.3 & 5.2/31.0 & \multicolumn{1}{c|}{6.1/26.8} & 62.3/14.1 & 55.8/11.1 & 55.0/11.2 & 55.1/6.3 & 50.5/7.2 \\ \hline
\multicolumn{11}{c}{\textbf{Likely to Interact (Correct/Incorrect)}} \\ \hline
\textbf{Definitely (C/I)} & 16.3/1.1 & 18.6/2.2 & 18/2.6 & 24.6/2 & \multicolumn{1}{c|}{18.9/3.5} & 0.7/15.3 & 0.8/18.5 & 0.6/12.4 & 0.0/15.4 & 0.9/18.2 \\
\textbf{Very Probably (C/I)} & 23.3/4.2 & 28.4/6.5 & 27.1/6.1 & 17.2/6.4 & \multicolumn{1}{c|}{27.4/5.0} & 3.4/15.3 & 5.4/11.8 & 5.6/17.2 & 7.7/23.1 & 6.5/18.7 \\
\textbf{Probably (C/I)} & 19.8/8.9 & 13.7/9.8 & 15.7/10.5 & 20.9/10.3 & \multicolumn{1}{c|}{15.2/11} & 2.7/13.9 & 7.8/21.0 & 7.9/16 & 7.7/13.8 & 7.5/19.1 \\
\textbf{Possibly (C/I)} & 17.4/12.6 & 21.6/17.9 & 20.3/17.6 & 17.9/20.6 & \multicolumn{1}{c|}{20.1/20.8} & 4.1/27.8 & 7.8/22.7 & 7.9/26.8 & 6.4/23.1 & 9.8/19.6 \\
\textbf{Probably Not (C/I)} & 16.3/26.3 & 9.8/23.4 & 11.8/24.4 & 11.9/24 & \multicolumn{1}{c|}{10.4/20.5} & 19.0/19.4 & 20.9/13.4 & 17.3/14.8 & 23.1/16.9 & 22.9/15.1 \\
\textbf{Definitely Not (C/I)} & 7.0/46.8 & 7.8/40.2 & 7.2/38.9 & 7.5/36.8 & \multicolumn{1}{c|}{7.9/39.1} & 70.1/8.3 & 57.4/12.6 & 60.7/12.8 & 55.1/7.7 & 52.3/9.3 \\ \hline
\end{tabular}
\caption{User's comfort levels and their likelihood to interact choices with different selected areas of interest.}
\label{tab:aoi-likely-comfort}
\vspace{-0.5mm}
\end{table*}

\subsubsection{Comparison of Comfortability and Interaction Rates Between Real and fake Messages.}

\hspace{0.4cm} Moving forward, we examine the impact of AOI on the comfort and interaction levels of participants. Given that comfortability and interaction scores are measured on a Likert scale, we conduct a T-test to compare the mean scores between real and fake SMS. For each message, we recorded participant comfort scores and likely interaction rates, analyzing their clicks to determine whether they fell within an AOI. The distribution of Likert choices made by participants is detailed in Table \ref{tab:aoi-likely-comfort}.

\textbf{Sender}
The 238 clicks placed on the sender AOI of fake messages (M = 2.26, SD = 1.56) compared with the 313 placed within real messages (M = 2.72, SD = 1.61) demonstrates significantly lower interaction rates with fake SMS (t(549) = 3.37, \textit{p} $<$ 0.001). The 238 clicks placed on the sender AOI of fake messages (M = 2.08, SD = 1.20) compared with the 313 placed within real messages (M = 2.49, SD = 1.28) demonstrate significantly lower comfort levels when interacting with fake SMS (t(523) = 3.86, \textit{p} $<$ 0.001).

\textbf{Entity}
There was no significant difference in interaction rates for participants who clicked on the Entity AOI when comparing real and fake messages. The 269 clicks placed on the sender AOI of fake messages (M = 2.47, SD = 1.27) compared with the 316 placed within real messages (M = 2.72, SD = 1.28) demonstrate significantly lower comfort levels when interacting with fake SMS (t(583) = 2.35, \textit{p} = 0.019).

\textbf{URL}
The 1096 clicks placed on the URL AOI of fake messages (M = 2.60, SD = 1.61) compared with the 660 placed within real messages (M = 2.9, SD = 1.61) demonstrates significantly lower interaction rates with fake SMS (t(1750) = 3.75, \textit{p} $<$ 0.001). 
The 1096 clicks placed on the URL AOI of fake messages (M = 2.08, SD = 1.20) compared with the 660 placed within real messages (M = 2.49, SD = 1.28) demonstrate significantly lower comfort levels when interacting with fake SMS (t(1754) = 5.77, \textit{p} $<$ 0.001).

\textbf{Call To Action} There was no significant difference in interaction rates for participants who clicked on the call to action AOI when comparing real and fake messages. The 166 clicks placed on the call to action AOI of fake messages (M = 2.45, SD = 1.24) compared with the 392 placed within real messages (M = 2.91, SD = 1.28) demonstrate significantly lower comfort levels with interacting with fake SMS (t(556) = 3.906, \textit{p} $<$ 0.001).

\textbf{Scenario}
There was no significant difference in interaction rates for participants who clicked on the scenario AOI when comparing real and fake messages. The 505 clicks placed on the scenario AOI of fake messages (M = 2.62, SD = 1.29) compared with the 531 placed within real messages (M = 2.80, SD = 1.24) demonstrate significantly lower comfort levels when interacting with fake SMS (t(1025) = 2.27, \textit{p} = 0.023).


\textbf{RQ3 Summary:} Our findings indicate that different Areas of Interest (AOIs) affected the participants' comfort levels and interaction rates with both real and fake messages differently. Specifically, for fake messages, identifying the sender and URL significantly reduced interaction rates and comfort levels. Conversely, for real messages, engaging with the call to action led to higher interaction rates and increased comfort levels. Furthermore, participants who incorrectly identified an AOI as influencing their comfort were significantly more likely to interact with fake messages. In contrast, those who accurately identified the sender were significantly less likely to engage with real messages.

\subsection{Answering RQ:4}

\hspace{0.4cm} We examined the correlation between post-survey results and user behavior, interaction rates, comfort levels, and accuracy in identifying real and fake messages using Spearman's rank correlation coefficient. These post-survey questions comprised Likert scales focused on privacy and security.

First, we examine the relationship between SA-6 scores and our accuracy and interaction rates. According to our Spearman's correlation analysis, we did not observe any significance between accuracy or interaction rates and SA-6 scores. 

Moving on, we investigate the relationship between IUIPC scores and our accuracy and interaction rates. After analyzing Spearman's correlation, we found a moderate, statistically significant decrease in interaction rates with fake messages with an increase in IUIPC scores (r$_{cor}$ = -0.337, \textit{p} $<$ 0.001). We also found a small, statistically significant decrease in interact rates with real messages with an increase in IUIPC scores (r$_{cor}$ = -0.260, \textit{p} $<$ 0.001). These findings suggest that participants who expressed higher concerns about their internet privacy were less likely to interact with both real and fake messages. However, no significant correlation was found between IUIPC scores and the overall message accuracy rates.

We then explore the relationship between RSeBIS scores and our accuracy and interaction metrics. Our Spearman's correlation revealed a small, statistically significant decrease in the interaction rates with fake messages as RSeBIS scores increased (r$_{cor}$ = -0.237, \textit{p} $<$ 0.001). These results indicate that participants who scored higher on the RSeBIS scale reported being less likely to interact with our fake messages. However, we found no significant correlations between RSeBIS scores and overall accuracy or interaction rates with real messages.

In our next post-survey test, we examined the correlation between familiarity with Security Terms and overall accuracy and interaction rates. Our analysis of the Spearman's correlation found a small, statistically significant decrease in interaction rates with fake messages with higher average Security Term Familiarity (r$_{cor}$ = -0.175, \textit{p} = 0.017). However, we found no significant correlations between Security Term Familiarity and overall accuracy or interaction rates with real messages.

Finally, we investigated the relationship between the Attention Control (ATTC) Scale and participants' accuracy and interaction rates. Our Spearman's correlation results show a small, statistically significant decrease in the interaction rates with both fake messages (r$_{cor}$ = -0.247, \textit{p} $<$ 0.001) and real messages (r$_{cor}$ = -0.238, \textit{p} $<$ 0.001) in accordance with a higher ATTC scale score. These results indicate that participants with a higher ATTC scale score were less likely to interact with either fake or real messages. However, we found no significant correlations between ATTC scale scores and overall accuracy.

\textbf{RQ4 Summary:} Our results indicate that participants scoring higher on the IUIPC, RSeBIS, Security Term Familiarity, and ATTC scales were less likely to engage with fraudulent messages. However, none of the behavioral scales significantly enhanced overall message accuracy. Consequently, individuals more concerned about privacy and exhibiting greater attention control appeared to be more skeptical towards all messages, leading them to be less comfortable and less likely to interact.
\section{Discussion}
\label{sec:discussion}

\subsection{Comparing with prior results.} 
In their work, Rahman et al.~\cite{rahman2023users} conducted a user study on smishing to investigate the rate at which users fall for smishing attacks. Their study reported that 16.92\% of participants fell for a smishing attack. In contrast, our study found that 34.4\% of fake messages were incorrectly identified as real. The difference in our rates can be attributed to several factors. In their study, successful smishing attacks were counted only when users actively interacted with the phishing messages. Thus, users who believed the messages were real but did not interact with them were not counted as being tricked by the attack. In our experiment, we created a controlled environment where users were asked to decide whether they believed the message was real or fake. Additionally, in this controlled environment, users did not benefit from potential built-in SMS protections or anti-phishing applications, such as security indicators in SMS apps that notify users when a message comes from an unknown number. Given these differences, we believe that our results offer a unique perspective on how users may be tricked by smishing.

\subsection{Areas of Interest and their effects on accuracy and message engagement.}
Previously, we identified the areas of interest that wielded the most significant influence on users' perceptions of real and fake messages. Our findings revealed that participants who identified areas of interest in real messages exhibited a significantly higher average comfort level with interaction and a higher likelihood of engaging compared to participants who identified similar areas of interest in fake messages. This discrepancy in comfort levels held true across all areas of interest. Notably, for likely interaction rates, significance was observed particularly in sender and URL clicks. It is crucial to emphasize the desired outcome: encouraging participants to engage more with real messages and less with fake ones. Furthermore, regarding the influence on participants' accuracy in the smishing detection task, we discovered that different areas of interest significantly impacted accuracy between real and fake messages.

In the case of real messages, pinpointing the entity and call to action markedly increased the likelihood of correctly identifying them. Interestingly, these areas are also the most susceptible to spoofing, as attackers can readily mimic genuine brand entities and the call to action found in legitimate messages. Nevertheless, paradoxically, these elements emerge as significant indicators that participants rely on to ascertain message authenticity.

In contrast, when it comes to fake messages, identifying the sender and URL significantly increases the chances of correctly detecting them. The sender and URL are crucial focal points for recognizing a fake message, as they are fundamental to the phishing method of attack. In phishing messages, users are lured into smishing by calling a number, replying to a text, or clicking on a link to visit a phishing website. Therefore, for that reply or call to direct to the attacker, it must be fraudulent, and the URL link must redirect to a fraudulent website for the attack to succeed. Additionally, while sender information can be spoofed, it's implicit that large businesses or federal institutions won't send you account updates or important notifications from 10 DLC numbers or email-to-text addresses. It is through heuristics like this that users can better identify fake messaging.

\subsection{More than 50\% of the time, users detect real SMS as fake.} 
In our research, we aimed to determine how the accuracy of user smishing detection across real and fake messages. Upon analyzing the results from our smishing detection task, we observed a consistent trend where participants were less accurate when identifying real messages compared to smishing messages. This resulted in an overall accuracy for real messages of 44.6\%, while the accuracy for fake messages was 65.6\%. Notably, this accuracy remained consistently at or below 50\% across all our attention score groups. Additionally, participants' post-survey results revealed that higher scores on the ATTC scale and IUIPC scores were negatively correlated with their interaction rates with real messages. Our hypothesis for this observation is that with the increased attention on real messages, users are finding more reasons to be concerned. This skepticism leads participants to be more cautious about interacting with all messages, regardless of whether they are fake or real.

\subsection{Recommendations}

\hspace{0.4cm}\textbf{Mobile Users need more education on short codes.} Our analysis revealed that while participants clicked 313 times on the sender in real messages, this led to (152/313) of those messages being incorrectly identified as fake and another (62/313) caused uncertainty. Notably, messages with short codes showed lower average comfort levels and likely interaction rates among participants. While short codes have been shown to be able to be spoofed, they are more difficult to spoof than regular phone numbers, and are generally considered safer~\cite{shortcodespoofing}. 

\textbf{How user training and app developers could improve user detection.}
Our findings can contribute to the design of improved educational tools and warning systems for detecting smishing attempts. By recognizing specific AOI that are often focus on or overlooked by users, developers of anti-smishing technologies can improve visibility of elements that lead to better recognition of smishing. This is supported by the findings that users with higher attention levels were more adept at correctly identifying smishing messages. In educational training, simulated scenarios akin to those utilized in this study could be created to underscore the significance of scrutinizing sender information and URLs in messages, emphasizing their pivotal role as primary indicators of a fake message. Furthermore, similar to how website developers can draw attention to particular aspects of web pages, mobile messaging app developers can enhance users' ability to identify fake messages by drawing their attention to these areas, which may make users less comfortable or inclined to interact with fake messages.

\subsection{Limitations}
\hspace{0.4cm}Our study has several limitations that warrant acknowledgment. Firstly, we note the disproportionate number of iOS screenshots compared to Android, with only two out of 16 screenshots being Android-based. This potential bias toward iOS users could impact the generalizability of our results, considering the influence of sampling bias on survey validity, as highlighted by Royal et al. \cite{royal2019survey}. However, we cannot definitively conclude whether participants' preference for a particular mobile OS significantly influenced their ability to detect smishing SMSes accurately in our survey.

Secondly, we acknowledge the potential compromise in results due to simulated study conditions \cite{schechter2007emperor, lain2022phishing}. Participants viewed screenshots within the Qualtrics survey, rather than directly from their messaging app. This approach was chosen to prioritize participant security and privacy. Additionally, it aimed to eliminate bias by providing with the same viewing context for each message. Lastly, it is important to recognize that in a setting where participants are aware of a security study being conducted, they tend to be more cautious. To mitigate this effect, we initially inquired about participants' likelihood to interact with and their comfort level in interacting with all messages before introducing the topic of real or fake messages at a later stage of the study.

Additionally, to measure to elements of the message content, we intentionally excluded the context of whether a message was expected or triggered by user activity. While this information can be important for distinguishing real from fake messages, we believe the necessary information to determine a message's authenticity can be found within the content itself. This approach allowed us to focus attention on the AOI in the message, acknowledging that this limitation may affect the realism of the study.

Lastly, we recognize the potential impact of survey fatigue, given the extensive nature of our survey. It comprised a pre-survey demographics questionnaire, 32 multi-part main survey questions, and five post-survey questionnaires. Fatigue may have affected participants' attention, potentially influencing result accuracy. Nevertheless, the substantial number of responses received suggests that participants engaged thoughtfully with our survey, and a high level of interaction can contribute to result validity \cite{wei2023skilled}.

\subsection{Future Work}
\hspace{0.4cm}
In a subsequent study, we aim to apply the principles established in this research to vulnerable populations, such as the elderly. Unfortunately, the elderly are more susceptible to online fraud compared to younger demographics \cite{FBI2022}. Individuals over the age of 60 experienced higher financial losses to fraud in 2022, losing a total of \$3.1 billion, which is 1.7 times more than the combined losses of individuals under the age of 20, 20-29, and 30-39, according to the 2022 FBI Elder Fraud report \cite{FBI2022}. 
Given these alarming losses, it is essential further to investigate the perceptions of the elderly towards smishing. This effort can help develop effective methods to protect them from the increasing threat of smishing attacks. Finally, based on our results we found significant differences between the likelihood and comfort in interacting with our smishing messages in various populations. Future work could explore potential socioeconomic factors that may contribute to this increase.

\section{Implications of Our Work}
\hspace{0.4cm}Our study unveils the effects that both human factors and message attributes have in smishing scenarios. Firstly, we establish the relationship between a user's perception of SMS messages and their ability to discern real from fake ones. This provides insights into the impact of different message attributes on a user's perception and identifies areas where participants may be misclassifying messages. Notably, our results indicate that users incorrectly predicted real messages as fake over 50\% of the time, suggesting that certain aspects of real messages trigger skepticism. Additionally, the results of our demographic analysis revealed that age, income level, and ethnicity were significant predictors in the user’s likelihood to fall for a smishing attack. Based on these findings, we anticipate a multitude of uses in anti-smishing technology and training.

In the development of anti-smishing technologies, it is crucial to comprehend the message attributes commonly linked with smishing to enhance their effectiveness in smishing detection and filtering tasks. This work has identified the aspects of SMS messages that users focused on for their decision, as well the areas which lead to the highest likelihood of successfully classifying messages. By taking both human factors and message attributes into consideration, a developer can create security warnings that are not only more noticeable to users but also more likely to impact user decision making. To accurately discern real from fake messages, users must focus on specific aspects of a message that can lead them to the correct answer and be familiar with the signs to look for. Despite participants focusing on aspects of messages that provide valuable information to the task of differentiating real and fake messages, our results show no significant improvement in overall accuracy across all message types. This highlights the need for smishing awareness programs, which would assist individuals in learning how to identify smishing attacks more effectively.

\section{Conclusion}
\label{sec:conclusion}

\hspace{0.4cm}
We conducted a smishing detection study involving 187 participants to unravel the factors influencing users' accuracy with both real and fake messages. Our findings indicate a smishing detection rate of 67.1\% for fake messages, contrasting with a lower 43.6\% for real SMSes. Additionally, we pinpointed areas of interest in messages that significantly affected users' accuracy in identifying both real and fake messages. These results underscore the considerable room for improvement in how brands communicate messages to their customers. Our findings will contribute to the development of warning systems for smishing and the creation of more effective educational tools aimed at increasing user awareness of identifying real and fake messages.

\bibliographystyle{plain}
\bibliography{main}

\section{Appendix}

\newpage
\onecolumn
\begin{figure}[H]
    \centering
    \includegraphics[clip, trim=6.5cm 10cm 8cm 10.5cm, width=.69\textwidth]{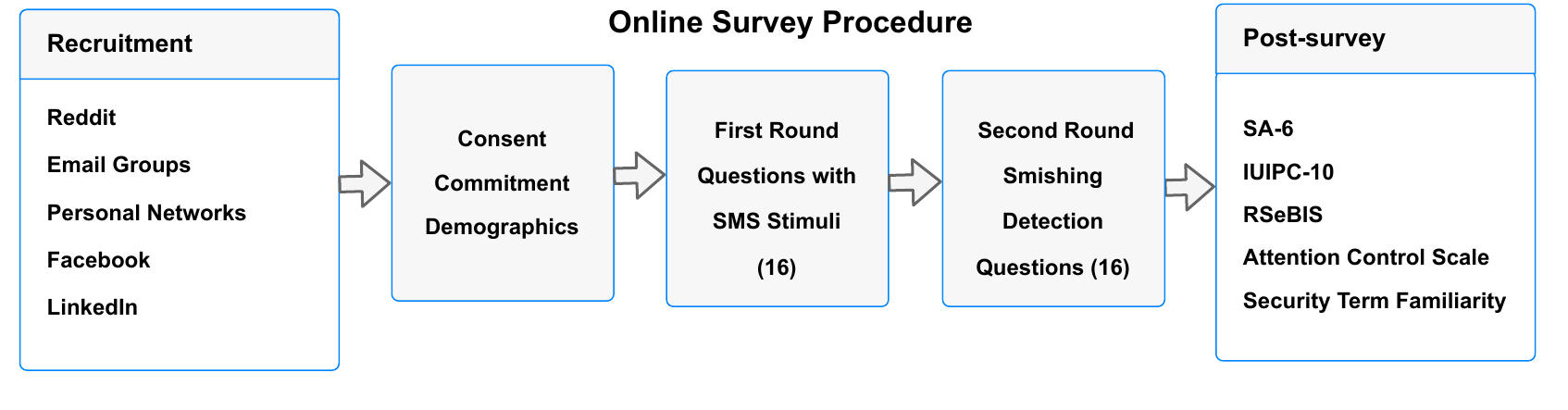}
    \caption{Overview of our online survey protocol.}
    \label{fig:mainsurvey}
    \vspace{-3mm}
\end{figure}

\begin{figure}[H]
    \centering    \includegraphics[width=.69\linewidth]{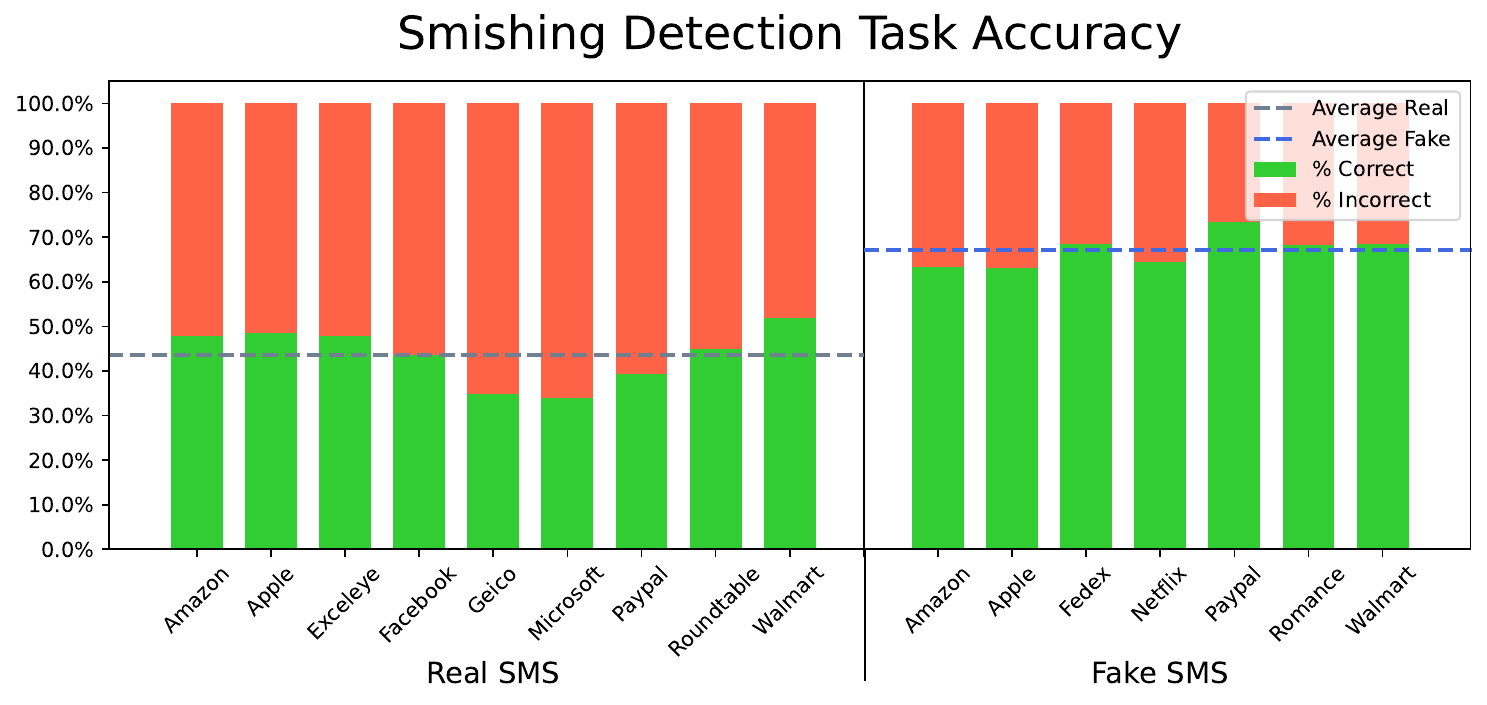}
    \caption{Accuracy for real and fake messages.}
    \label{tab:accuracy-rates-bymessage}
\end{figure}

\begin{table}[H]
\centering
\begin{tabular}{llccccccc}
\hline
\multirow{2}{*}{Cat.} & \multicolumn{1}{c}{\multirow{2}{*}{Range}} & \multirow{2}{*}{N(\%)} & \multicolumn{2}{c}{Likely Rates} & \multicolumn{2}{c}{Comfort Levels} & \multicolumn{2}{c}{Acc \%} \\ \cline{4-9} 
 & \multicolumn{1}{c}{} &  & \multicolumn{1}{c|}{Real} & \multicolumn{1}{c|}{Fake} & \multicolumn{1}{c|}{Real} & \multicolumn{1}{c|}{Fake} & Real & \multicolumn{1}{c|}{Fake} \\ \hline
\multirow{3}{*}{\rotatebox[origin=c]{90}{Sex}} & Female & 81(43.3) & 2.69 & 2.49 & 2.61 & 2.28 & 40.0 & 74.0 \\
 & Male & 103(55.1) & 2.95 & 2.90 & 2.73 & 2.62 & 47.4 & 60.5 \\
 & Other/Prefer not to say & 3(1.6) & 2.10 & 1.62 & 2.28 & 1.86 & 57.1 & 50.0 \\ \hline
\multirow{7}{*}{\rotatebox[origin=c]{90}{Ethnicity}} & White & 119(63.6) & 2.91 & 2.85 & 2.79 & 2.62 & 46.1 & 64.4 \\
 & Asian & 13(7.0) & 2.11 & 1.89 & 2.12 & 2.04 & 26.6 & 82.5 \\
 & Black & 14(7.5) & 3.20 & 3.15 & 2.94 & 2.63 & 42.1 & 54.1 \\
 & Hispanic or Latino & 10(5.3) & 1.86 & 1.40 & 1.82 & 1.31 & 24.9 & 93.2 \\
 & Native American & 19(10.2) & 3.15 & 3.09 & 2.81 & 2.76 & 58.3 & 53.1 \\
 & Mixed or Biracial & 8(4.3) & 2.63 & 2.01 & 2.24 & 1.53 & 52.5 & 73.4 \\
 & Prefer not to say & 4(2.1) & 2.63 & 2.11 & 2.30 & 1.95 & 12.5 & 80.0 \\ \hline
\multirow{5}{*}{\rotatebox[origin=c]{90}{Age}} & 18 - 24 & 26(13.9) & 2.32 & 1.98 & 2.35 & 1.90 & 28.0 & 77.9 \\
 & 25 - 34 & 116(62.0) & 2.78 & 2.63 & 2.62 & 2.43 & 46.5 & 67.9 \\
 & 35 - 44 & 35(18.7) & 3.14 & 3.27 & 2.93 & 2.85 & 47.7 & 53.9 \\
 & 45 - 54 & 5(2.7) & 3.06 & 3.03 & 3.09 & 2.82 & 74.6 & 40.8 \\
 & 55+ & 5(2.7) & 4.09 & 3.83 & 3.47 & 3.23 & 42.2 & 61.4 \\ \hline
\multirow{8}{*}{\rotatebox[origin=c]{90}{Income}} & Less than \$10,000 & 5(2.7) & 1.42 & 1.11 & 2.00 & 1.71 & 16.7 & 96.4 \\
 & 10, 000-19,999 & 10(5.3) & 2.82 & 2.61 & 1.96 & 1.98 & 24.9 & 80.4 \\
 & 20, 000-39,999 & 19(10.2) & 3.23 & 3.17 & 2.89 & 2.78 & 47.5 & 58.7 \\
 & 40, 000-59,999 & 38(20.3) & 3.06 & 3.12 & 2.92 & 2.80 & 44.9 & 60.9 \\
 & 60, 000-79,999 & 33(17.6) & 2.67 & 2.53 & 2.48 & 2.29 & 42.0 & 70.9 \\
 & 80, 000-99,999 & 43(23.0) & 2.88 & 3.03 & 2.80 & 2.76 & 50.4 & 54.2 \\
 & \$100,000 or more & 27(14.4) & 3.01 & 2.40 & 2.91 & 2.25 & 52.9 & 75.9 \\
 & Prefer not to say & 12(6.4) & 1.81 & 1.37 & 1.89 & 1.44 & 25.0 & 80.3 \\ \hline
\multirow{5}{*}{\rotatebox[origin=c]{90}{Phone Use}} & \textgreater{}1 - \textless{}2 hours & 5(2.7) & 3.01 & 2.03 & 2.64 & 2.39 & 56.3 & 58.3 \\
 & \textgreater{}2 - \textless{}3 hours & 31(16.6) & 2.69 & 2.77 & 2.64 & 2.48 & 41.0 & 64.9 \\
 & \textgreater{}3 - \textless{}4 hours & 53(28.3) & 3.00 & 2.83 & 2.81 & 2.56 & 41.1 & 64.4 \\
 & \textgreater{}4 - \textless{}5 hours & 28(15.0) & 2.76 & 2.62 & 2.59 & 2.38 & 49.8 & 69.9 \\
 & \textgreater{}5 hours & 70(37.4) & 2.78 & 2.65 & 2.62 & 2.42 & 45.0 & 66.6 \\ \hline
\multirow{9}{*}{\rotatebox[origin=c]{90}{Education}} & Less than high school & 2(1.1) & 4.17 & 4.36 & 3.67 & 3.50 & 83.3 & 33.3 \\
 & High school graduate & 17(9.1) & 2.64 & 2.57 & 2.58 & 2.44 & 32.7 & 62.3 \\
 & Some college & 28(15.0) & 2.86 & 2.57 & 2.81 & 2.47 & 52.9 & 68.4 \\
 & Associate degree & 19(10.2) & 2.35 & 2.30 & 2.44 & 2.11 & 31.5 & 75.4 \\
 & Bachelor’s degree & 73(39.0) & 2.90 & 2.81 & 2.73 & 2.49 & 45.6 & 65.8 \\
 & Master’s degree & 33(17.6) & 2.79 & 2.74 & 2.50 & 2.45 & 44.1 & 61.8 \\
 & Doctoral degree & 7(3.7) & 3.48 & 3.06 & 2.92 & 2.73 & 46.1 & 61.4 \\
 & Professional degree & 5(2.7) & 2.40 & 2.05 & 2.53 & 2.60 & 51.3 & 77.1 \\
 & Other/Prefer not to say & 3(1.6) & 3.94 & 3.24 & 3.00 & 2.93 & - & 100 \\ \hline
\multirow{12}{*}{\rotatebox[origin=c]{90}{Occupation}} & Administrative support & 13(7.0) & 2.71 & 2.45 & 2.71 & 2.34 & 46.3 & 70.8 \\
 & Art/Writing/Journalism & 16(8.6) & 2.91 & 2.76 & 2.61 & 2.43 & 42.4 & 61.8 \\
 & Bus./Mgmt./Fin. & 34(18.2) & 2.96 & 3.00 & 2.82 & 2.63 & 50.4 & 63.1 \\
 & Education or Science & 18(9.6) & 2.81 & 2.76 & 2.74 & 2.60 & 48.5 & 61.7 \\
 & Legal & 6(3.2) & 2.87 & 2.45 & 2.69 & 2.36 & 74.5 & 41.5 \\
 & Medical & 8(4.3) & 2.98 & 3.24 & 2.59 & 2.91 & 33.3 & 67.9 \\
 & IT & 21(11.2) & 3.12 & 3.16 & 2.91 & 2.86 & 45.9 & 54.2 \\
 & Engineer(other) & 13(7.0) & 2.76 & 2.63 & 2.53 & 2.28 & 25.7 & 86.6 \\
 & Service & 13(7.0) & 3.12 & 3.14 & 2.90 & 2.89 & 61.1 & 60.6 \\
 & Skilled Labor & 7(3.7) & 2.48 & 2.69 & 2.35 & 2.42 & 32.3 & 51.4 \\
 & Student & 18(9.6) & 2.36 & 1.69 & 2.45 & 1.79 & 31.6 & 85.7 \\
 & Other/Prefer not to say & 20(10.7) & 2.64 & 2.30 & 2.44 & 2.06 & 46.1 & 70.0 \\ \hline
\end{tabular}
\caption{Demographic variables by average likely to interact, comfort levels and overall accuracy rates.}
\label{tab:DemographicAttribute2}
\end{table}

\onecolumn

\begin{table*}[htbp]
  \centering
  \tiny
  \setlength{\tabcolsep}{5pt}
  \renewcommand{\arraystretch}{1.5}
\begin{tabular}{llll}
\hline
\textbf{Phish?} & \textbf{Category} & \textbf{Sender} & \textbf{SMS Body} \\ \hline
Phish & Email to Text & \begin{tabular}[c]{@{}l@{}}myappid6583063.flixid \\ numbr59022.int19440me@...\end{tabular} & \begin{tabular}[c]{@{}l@{}}qu2zAOGOys \#{[}Netflix Subscription Plans{]} We could not successfully process your payment, and \\ your subscription will remain active until May 14, 2023. It's quick and easy to restart your membership by clicking the secure link below: \\ \url{https://app.atlcunion.org/a0c418} Come back and enjoy newly added popular movies and full seasons of hit TV shows. \\ Thanks for choosing Netflix. -The Netflix Team\end{tabular} \\ \hline
Phish & Email to Text & 1-888-280-0835.inc@vip-smsinc.... & \begin{tabular}[c]{@{}l@{}}Notification From Amazon: We've locked your Amazon account due to a billing issue. \\ To unlock your Amazon account, please click the link below: \url{https://qr.io/r/GJ1jvX?amazon-billing-update \& ?=https://amazon.com} \\ Please take action on your account within 48 hours to avoid permanent suspension. Regards, Amazon Service\end{tabular} \\ \hline
Phish & Email to Text & notification-noreply-07-apple.co... & \begin{tabular}[c]{@{}l@{}}Your Apple ID has been locked. We have locked your Apple ID because our service has detected two unauthorized devices. \\ To unlock your account, you are required to verify your Apple ID. Click the link below to unlock your Apple ID. \\ \url{http://s948917531.onlinehome.us/} Your account will be automatically unlocked after finishing the verification. \\ Copyright © 2023 Apple Distribution International, Hollyhill Industrial Estate, Hollyhill, Cork, Irlande. All rights reserved.\end{tabular} \\ \hline
Phish & Number & 1410100014 & \begin{tabular}[c]{@{}l@{}}FRM:719132-Fed\_EX.906697 MSG:flm-;it will be returned to sender, We have made several attempts to reach you, \\ https://jf245-fedex.me/yjDxio?37588372\end{tabular} \\ \hline
Real & Short code & 25392 & \begin{tabular}[c]{@{}l@{}}\url{https://amazon.com/a/c/r/YyRZTFC7nfNMAUY7iB1FfPPAW} Amazon: Sign-in attempt from CA, US. \\ Tap the link to respond.\end{tabular} \\ \hline
Real & Short code & 20993 & \begin{tabular}[c]{@{}l@{}}Hi, this is Apple Support. Thanks for agreeing to take our short survey. To opt out of this survey, text STOP. \\ Please go to \url{https://s.apple.com/Bb8V3D4d20} \end{tabular} \\ \hline
Real & Short code & 673804 & Microsoft: Password changed for *********60. Not you? \url{https://aka.ms/alcr} \\ \hline
Real & Short code & 43426 & \begin{tabular}[c]{@{}l@{}}GEICO Policy: Renewal ID Cards are now available at \url{https://geico.app.link/smsExpressRenl} for your auto policy ending in 2132.\\  Reply STOP to end texts.\end{tabular} \\ \hline
Real & Short code & 32665 & jeff, you have 24 new notifications on Facebook: \url{https://fb.com/l/2E7aH1P7DLOLsJ6} \\ \hline
Phish & Email to Text & support@722-paypal3125-30069.com & \begin{tabular}[c]{@{}l@{}}(PayPal-Issue: Your account has been restricted. Check it here Immediately. \textgreater \\ \url{https://me2.do/GBIOWzEt?V4ZGKR} $\sim$) 29FNO\end{tabular} \\ \hline
Phish & Number & +1 (951) 923-3865 & \begin{tabular}[c]{@{}l@{}}1000 Congrats BEN! Your code 9FR-S3R7 printed on your last receipt is among 7 we randomly picked for \$1000 \\ Walmart gift card promotion \url{ab4nr.xyz/S1yrXsApa} \end{tabular} \\ \hline
Phish & Number & +1 (631) 739-5714 & Mr. Williams, this is Laura. We had a long chat on the dating website a week ago. I hope my messages don't bother you. \\ \hline
Real & Short code & 729-725 & PayPal: For assistance, please visit the Help Center \url{https://www.paypal.com/help} \\ \hline
Real & Short code & 61746 & Your Walmart package is on the way! Track it in your order details: \url{https://w-mt.co/g/6ytlzY} Reply HELP for info; STOP to opt out \\ \hline
Real & Short code & 34185 & \begin{tabular}[c]{@{}l@{}}Hi Daniel, it's time to schedule your next Annual Eye Exam with Excel Eyecare Optometry. Please call 8587809889 or click \\ \url{4pc.me/dM5YG3hpzhb} \end{tabular} \\ \hline
Real & Short code & 866-77 & \begin{tabular}[c]{@{}l@{}}Make the last pizza of the year a Round Table pizza! Get \$7 off an L or XL pizza today. Get your code: \\ \url{https://mfon.us/ck672aev72r} HELP/STOP call 8447887525\end{tabular} \\ \hline
\end{tabular}
  \caption{List of SMSes being used in our study.}
  \label{fig:msglist}
\end{table*}

\onecolumn

\begin{figure}[htbp]
    \centering
    \begin{subfigure}{0.19\textwidth}
        \centering
        \includegraphics[height=5.3cm]{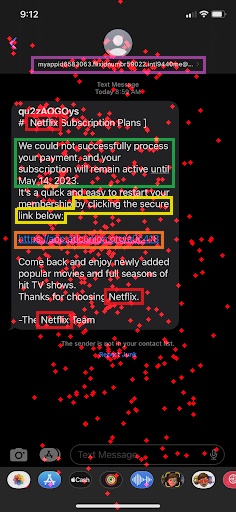}
        \caption{\\Netflix}
    \end{subfigure}%
    \begin{subfigure}{0.19\textwidth}
        \centering
        \includegraphics[height=5.3cm]{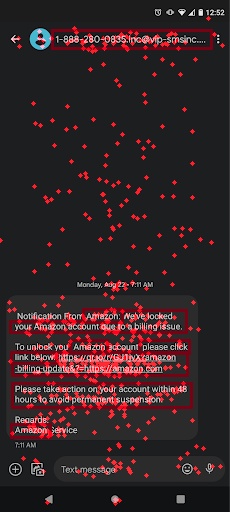}
        \caption{\\Amazon}
    \end{subfigure}%
    \begin{subfigure}{0.19\textwidth}
        \centering
        \includegraphics[height=5.3cm]{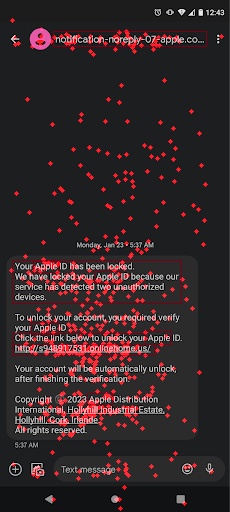}
        \caption{\\Apple}
    \end{subfigure}
    \begin{subfigure}{0.19\textwidth}
        \centering
        \includegraphics[height=5.3cm]{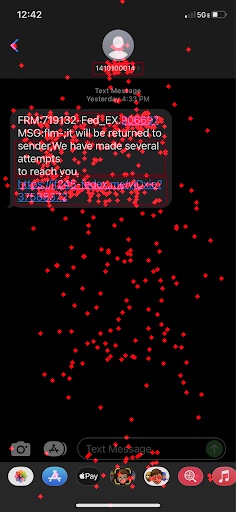}
        \caption{\\FedEx}
    \end{subfigure}
    \begin{subfigure}{0.19\textwidth}
        \centering
        \includegraphics[height=5.3cm]{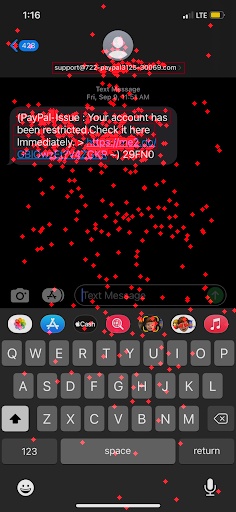}
        \caption{\\PayPal}
    \end{subfigure}\\
    \begin{subfigure}{0.19\textwidth}
        \centering
        \includegraphics[height=5.3cm]{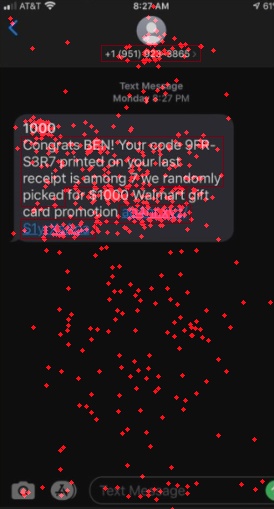}
        \caption{\\Walmart}
    \end{subfigure}%
    \begin{subfigure}{0.19\textwidth}
        \centering
        \includegraphics[height=5.3cm]{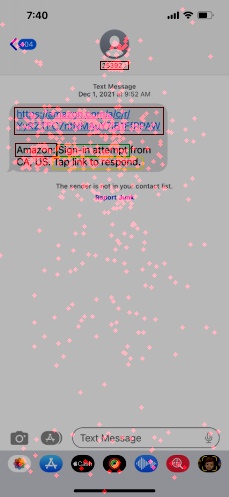}
        \caption{\\Amazon}
    \end{subfigure}
    \begin{subfigure}{0.19\textwidth}
        \centering
        \includegraphics[height=5.3cm]{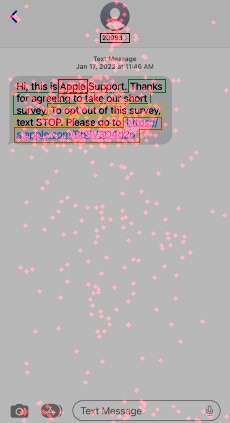}
        \caption{\\Apple}
    \end{subfigure}
    \begin{subfigure}{0.19\textwidth}
        \centering
        \includegraphics[height=5.3cm]{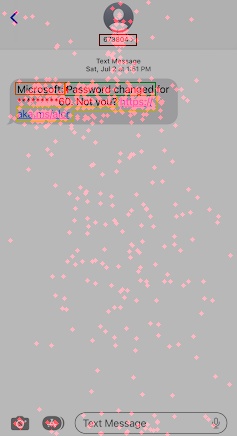}
        \caption{\\Microsoft}
    \end{subfigure}
    \begin{subfigure}{0.19\textwidth}
        \centering
        \includegraphics[height=5.3cm]{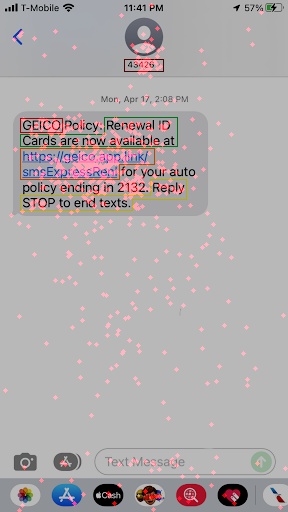}
        \caption{\\Geico}
    \end{subfigure}\\
    \begin{subfigure}{0.19\textwidth}
        \centering
        \includegraphics[height=5.3cm]{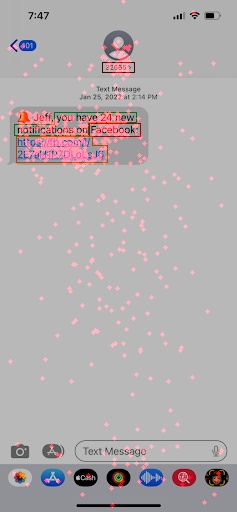}
        \caption{\\Facebook}
    \end{subfigure}
    \begin{subfigure}{0.19\textwidth}
        \centering
        \includegraphics[height=5.3cm]{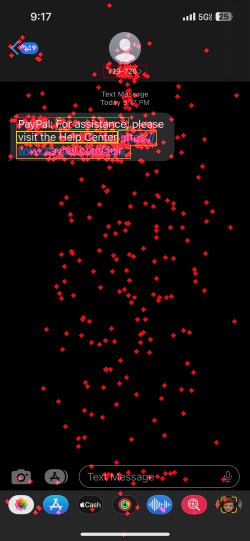}
        \caption{\\PayPal}
    \end{subfigure}
    \begin{subfigure}{0.19\textwidth}
        \centering
        \includegraphics[height=5.3cm]{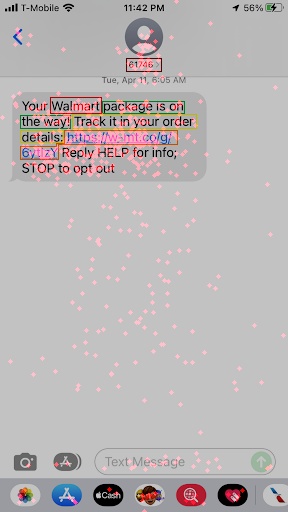}
        \caption{\\Walmart}
    \end{subfigure}
    \begin{subfigure}{0.19\textwidth}
        \centering
        \includegraphics[height=5.3cm]{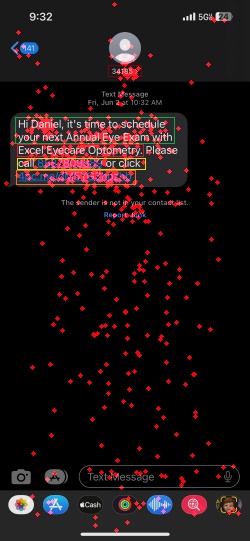}
        \caption{\\Excel Eye Care}
    \end{subfigure}
    \begin{subfigure}{0.19\textwidth}
        \centering
        \includegraphics[height=5.3cm]{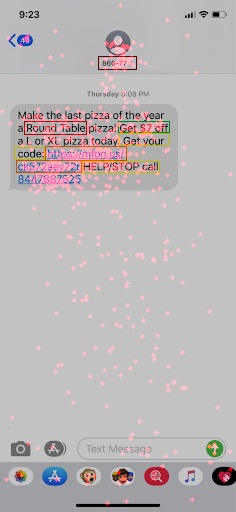}
        \caption{\\Round Table}
    \end{subfigure} \\
    \caption{Messages used in experiment marked with areas of interest and click points. The first six screenshots are fake, the remaining nine are real.}
    \label{fig:aoi_images}
\end{figure}

\end{document}